\documentclass[%
 reprint,
superscriptaddress,
 amsmath,amssymb,
 aps,
onecolumn,
showkeys,
]{revtex4-2}
\usepackage{graphicx}
\usepackage{dcolumn}
\usepackage{bm}
\usepackage{hyperref}
\usepackage{multirow}
\usepackage{adjustbox}
\usepackage{graphics}
\usepackage{subfig}

\begin{document}

\title{Properties of $\Omega_{ccb}$ and $\Omega_{bbc}$ baryons in a quark–diquark model}

\author{Chaitanya Anil Bokade}
\email{anshul497@gmail.com}
\altaffiliation[ORCID iD: ]{0009-0007-5463-6812}
\author{Bhaghyesh}%
\email{bhaghyesh.mit@manipal.edu; Corresponding author}
 \altaffiliation[ORCID iD: ]{0000-0003-3994-9945}
\affiliation{Manipal Institute of Technology\\Manipal Academy of Higher Education, Manipal, 576104, Karnataka, India}

\begin{abstract}
In this work, we investigate the mass spectra and radiative transitions of $\Omega_{ccb}$ and $\Omega_{bbc}$ baryons within a relativistic screened potential model using the quark–diquark approximation to capture key aspects of their internal structure. The model describes the $S$-, $P$-, and $D$-wave excitations, provides corresponding mass predictions, and offers a comparison with other theoretical approaches. Our findings suggest that excitations involving the diquark occur at lower energies than those requiring relative motion between the quark and diquark, challenging conventional assumptions about the rigidity of heavy diquark systems. Analysis of radiative decays of these states, shows that spin–flip transitions within the same orbital multiplet are suppressed, while transitions involving orbital or radial changes dominate with significantly larger widths. The model offers subtle distinctions in excitation dynamics and highlights its value for future studies and for interpreting potentially upcoming experimental results on triply heavy baryons.


\end{abstract}

\maketitle

\newpage

\section{\label{sec:Introduction} Introduction}

Heavy baryon spectroscopy is crucial for improving our understanding of the strong interaction within the non-perturbative regime of Quantum Chromodynamics (QCD), the foundational theory that describes quarks and gluons. Over the last five decades, QCD has become a cornerstone of contemporary particle physics, with one of its fundamental predictions being the presence of heavy hadrons, which are composite particles comprising one or more heavy quarks. The field has sparked significant theoretical and experimental interest, particularly in the last two decades, with various experimental groups such as LHCb, CLEO, BABAR, CDF, BELLE, and BESIII discovering a diverse range of heavy-flavour hadrons, including heavy mesons, singly and doubly heavy baryons, and exotic multiquark states such as hidden-charm tetraquarks and pentaquarks \cite{1}. The exploration of baryons with single, double, or even triple heavy quarks provides fundamental insights into various phenomena such as heavy quark symmetry, chiral dynamics, and the structure of strong interactions in the non-perturbative QCD domain. The most recent Particle Data Group (PDG) review lists more than $70$ experimentally established singly heavy baryons \cite{1}. The successful identification of nearly all ground-state heavy baryons has shifted much of the theoretical work is focused on single heavy baryons \cite{2,3}. Experimental research on doubly heavy baryons has witnessed remarkable progress in recent years, with a number of notable achievements. A major milestone was achieved in 2017 when the LHCb Collaboration announced the first definitive discovery of the doubly charmed baryon $\Xi^{++}_{cc}(3621)$ in the $\Lambda^{+}_{c}K^{-}\pi^{+}\pi^{+}$ mass spectrum \cite{4}, which is 100 MeV higher from the $\Xi^{+}_{cc}(3520)$ state observed by SELEX collaboration \cite{5,6}. Following the discovery, experimentalists measured the lifetime of $\Xi^{++}_{cc}$ \cite{7}, providing crucial input for theoretical models of heavy quark dynamics. Additionally, two new decay channels, $\Xi^{++}_{cc} \rightarrow \Xi^{+}_{c}\pi^{+}$ and $\Xi^{++}_{cc} \rightarrow \Xi'^{+}_{c}\pi^{+}$ were successfully identified \cite{8,9}, significantly improving our understanding of its decay properties. Triply heavy baryons, which are composed entirely of heavy charm and/or bottom quarks, are the last remaining members of the standard heavy baryon family that have yet to be discovered experimentally. Their production presents substantial challenges due to the demand for greater energies and the naturally low production rates associated with such massive systems \cite{10,11,12,13}. It has been argued that the likelihood of producing triply heavy baryons in $e^{+}e^{-}$ collisions is extremely limited \cite{14,15}. Optimistic projections suggest that the LHC might play a critical role in their discovery. For instance, estimates in Ref. \cite{16} indicate that $10^{4}-10^{5}$ events of triply heavy baryons with $ccc$ and $ccb$ content could be recorded for an integrated luminosity of $10\text{ fb}^{-1}$ at the LHC. Theoretical studies have also suggested alternative search techniques, such as concentrating on semi-leptonic and non-leptonic decay modes \cite{17,18,19}, as well as examining their production in the quark-gluon plasma (QGP) setting of relativistic heavy-ion collisions, where several studies have indicated that the yields of fully heavy baryons, such as $\Xi_{cc}$ and $\Omega_{ccc}$,could be significantly increased in heavy-ion collisions at RHIC and LHC compared to proton-proton collisions in vacuum \cite{13,20,21,22,23,24}. The pursuit of triply heavy baryons has therefore gained prominence, and it has been identified as one of the primary priorities for future heavy-ion studies at the LHC \cite{25}. From a theoretical standpoint, triply heavy baryons provide a unique platform for studying heavy quark dynamics without the complications introduced by light quark interactions, resulting in cleaner conditions to test QCD in the fully heavy sector and possibly revealing novel aspects of baryon structure. Over the years, numerous theoretical frameworks have been employed to predict the mass spectra and other properties of these triply heavy baryons, such as, the Bag Model \cite{26,27}, Non-relativistic Constituent Quark Models (NRCQM) \cite{28,29,30}, the Diquark Model \cite{31,32}, QCD Sum Rules \cite{33,34,35,36,37}, Lattice QCD \cite{38,39}, Regge Theory \cite{40}, Non-relativistic QCD (NRQCD) \cite{41}, the Faddeev Approach \cite{42,43,44}, the Hyper-central Quark Model (HCQM) \cite{45,46,47}, Variational Methods \cite{48,49}, the Relativistic Quark Model (RQM) \cite{50,51}, the Renormalization Group Procedure for Effective Particles (RGPEP) \cite{52}, etc.

Among various systematic studies, Ref. \cite{29} is one of the earliest. It presents a thorough analysis utilizing a non-relativistic quark model to construct a unified theoretical framework for baryons, including those with heavy quarks, by solving the three-body problem explicitly. Most theoretical work on triply heavy baryons has focused on calculating the mass spectra of ground and low-lying excited states. Since these baryons are composed entirely of heavy quarks, they are often treated as non-relativistic systems in majority of the literature \cite{29,30,45,46}. However, a comprehensive understanding requires the investigation of higher radial and orbital excitations as well. Exploring the full spectrum can provide deeper insights into the underlying structure and interactions. One recurring challenge in baryon spectroscopy is the discrepancy between the number of predicted excited states and those observed experimentally. This difference, which initially appeared in the context of light-quark baryons, is known as the “missing resonance” problem \cite{53}. A proposed resolution involves reducing the internal degrees of freedom by considering a quark–diquark structure, in which the baryon is represented as a bound state of a single quark and a compact diquark. This reduces the number of internal degrees of freedom and naturally leads to a sparser spectrum of excitations \cite{54}. Triply heavy baryons can be treated either through the quark–diquark approximation or by directly solving the full three-body Schrödinger equation \cite{30,32,50}. Although the three-body approach is often viewed as more fundamental, there is no conclusive evidence regarding the internal dynamics of quarks within baryons. Furthermore, lattice QCD simulations support a Y-shaped string configuration, which is compatible with the quark–diquark model \cite{55}. Additionally, the quark–diquark picture has shown notable success in reproducing the spectra of heavy-light baryons \cite{3}, lending further support to its use. Assuming this quark–diquark structure, the energy of a heavy diquark is expected to scale as $\alpha_{s}^{2}(M_{Q})M_{Q}$, implying that excitations of the heavy diquark should have higher energy than those associated with light degrees of freedom \cite{56}. However, several theoretical studies contradict this observation \cite{29,32,51,57,58,59}. For instance, Ref. \cite{29}, suggests that when the heavy quark mass grows, so does the energy differences between states with different kinds of excitations, implying that such excitations may be more accessible than previously thought. To investigate such excitations rigorously, one must carefully consider the dynamics of two heavy quarks, for which Non-relativistic QCD (NRQCD)  \cite{60}, provides an appropriate theoretical framework. NRQCD has since been evolved into two effective field theories: potential NRQCD (pNRQCD) and velocity NRQCD (vNRQCD). In pNRQCD, soft gluon modes are integrated out, making it possible to describe bound heavy diquark states with fixed color structures \cite{61}. In contrast, vNRQCD keeps individual heavy quarks as active degrees of freedom and matches the theory at the hard scale \cite{62}. Although compact diquarks are expected in the infinite-mass limit, they may also form for moderately heavy quarks if the confinement mechanism contains significant contributions beyond the Coulomb interaction \cite{63}. Charm quarks, for instance, are not heavy enough to be fully in the Coulombic regime and thus may be subject to additional nonperturbative effects. These effects, though not systematically included in traditional NRQCD, may support an approximate doubly heavy diquark–antiquark (DHDA) symmetry \cite{63}. This suggests the need for a new theoretical framework that consistently integrates both perturbative and nonperturbative contributions, possibly by combining aspects of NRQCD and heavy quark effective theory (HQET) \cite{63}. Introducing confining potentials could effectively reduce the diquark size, making the application of diquark-based approximations more viable \cite{63}. In systems involving heavy quarks, such as charm and bottom, a clear hierarchy of energy scales emerges: $m \gg mv \gg mv^{2}$ \cite{64}. This allows one to integrate out higher-momentum modes, moving from complete QCD to its nonrelativistic forms such as NRQCD and pNRQCD \cite{64}. By further neglecting color transitions between singlet and octet states, pNRQCD reduces into a potential model, wherein the Schrödinger equation can be used to describe the system \cite{47,64}. The Cornell potential in such models has been highly successful in explaining quarkonium spectra \cite{65}. The Schrödinger equation has also been extended to describe three-body systems, resulting in predictions for the masses of triply heavy baryons \cite{28,50,30,65,66}. Nonetheless, studies of heavy quarkonium \cite{65} have highlighted the significance of relativistic corrections, especially in systems involving charm quarks. To incorporate these effects, the relativized quark model was developed by \cite{67,68}, providing a more refined framework. This approach allows for the treatment of relativistic effects while preserving much of the intuitive structure of non-relativistic models. Building on the insights from the above literature, we adopt our previously developed relativistic screened potential model \cite{69,70,71} and extend its application to the study of triply heavy baryons. Following our earlier study of the $\Omega_{ccc}$ and $\Omega_{bbb}$ baryons within the quark–diquark framework \cite{71a}, we now investigate the mass spectra and radiative decay properties of the $\Omega_{ccb}$ and $\Omega_{bbc}$ baryons.

In this article, we present a comprehensive analysis of the $\Omega_{ccb}$ and $\Omega_{bbc}$ baryons within the framework of a relativistic screened potential quark–diquark model. Section \ref{sec:Methodology} outlines the theoretical formalism employed to describe the bound-state dynamics of these baryons, along with the numerical methods used to solve the relativistic Schrödinger equation. In Section \ref{sec:Radiative Transitions}, we examine the radiative decay processes associated with these systems. Section \ref{sec:Results and Discussion} provides an in-depth analysis of our results, including a comparative discussion with other theoretical models. Finally, Section \ref{sec:Conclusion} summarizes our findings and conclusions.

\section{\label{sec:Methodology} Methodology}

To compute the mass spectra of the $\Omega_{ccb}$ and $\Omega_{bbc}$ baryons within the quark-diquark framework, we treat the diquark as consisting of two quarks of the same flavor. The computation of baryon spectra is done in two steps: first, the mass of the heavy diquark is calculated; this mass is then utilized as input to model the baryon as a bound state of a quark and a diquark. To account for relativistic effects in kinematics, we have used the relativistic generalization of the non-relativistic Hamiltonian given by \cite{67,68}

\begin{equation}
	\label{eq:1}
	H = \sqrt{-\nabla_{1}^{2} + m_{1}^{2}}+\sqrt{-\nabla_{2}^{2} + m_{2}^{2}} + V(r) \,.
\end{equation}

\noindent Here, $\vec{r} = \vec{x}_{1} - \vec{x}_{2}$, where $\vec{x}_{1}$, $m_{1}$ and $\vec{x}_{2}$, $m_{2}$ denote the coordinates and masses of the two constituent quarks in the case of a diquark, or the quark and diquark in the case of a baryon. The operators $\nabla_{1}^{2}$ and $\nabla_{2}^{2}$ represent the Laplacians with respect to the coordinates $\vec{x}_{1}$ and $\vec{x}_{2}$, respectively. To obtain the masses, the Schrodinger equation is solved as an eigenvalue equation using the method developed in Refs. \cite{72,73}. A brief summary of the method is given below. The relativistic wave equation corresponding to the above Hamiltonian is given by

\begin{equation}
	\left(\sqrt{-\nabla_{1}^{2} + m_{1}^{2}}+\sqrt{-\nabla_{2}^{2} + m_{2}^{2}} + V(r)\right)\Psi(\vec{r})  = E\Psi(\vec{r}) \label{waveeq} \,.
\end{equation}

\noindent Expanding the wave function in terms of spectral integration,
\begin{equation}
	\Psi(\vec{r}) = \int d^{3}r'\int \frac{d^{3}k}{(2\pi)^{3}}e^{i\vec{k}(\vec{r}-\vec{r}')}\Psi(\vec{r}') \,,
\end{equation}

\noindent we can rewrite Eq.\eqref{waveeq} as

\begin{widetext}
\begin{equation}
	\label{eq:2}
	V(r)\Psi\left(\vec{r}\right)+\int d^{3}r'\frac{d^{3}k}{(2\pi)^{3}}\left(\sqrt{k^{2} + m_{1}^{2}}+\sqrt{k^{2} + m_{2}^{2}}\right)e^{i\vec{k}(\vec{r}-\vec{r}')}\Psi\left(\vec{r}'\right) = E\Psi\left(\vec{r}\right) \,.
\end{equation}
\end{widetext}

\noindent The exponential term in the above equation can be decomposed in terms of spherical harmonics as

\begin{equation}
	e^{i\vec{k}\cdot\vec{r}} = 4\pi\sum_{nl}Y_{nl}^{*}\left(\hat{k}\right)Y_{nl}(\hat{r})j_{l}(kr)i^{l} \label{expterm} \,,
\end{equation}

\noindent where $j_{l}$ is the spherical Bessel function, $Y_{nl}^{*}(\hat{k})$ and $Y_{nl}(\hat{r})$ are the spherical harmonics with normalization condition $\int d\Omega Y_{n_{1}l_{1}}(\hat{k})Y_{n_{2}l_{2}}(\hat{r}) = \delta_{n_{1}n_{2}}\delta_{l_{1}l_{2}}$, $\hat{k}$ and $\hat{r}$ are unit vectors along the $\vec{k}$ and $\vec{r}$ direction, respectively. Factoring the wave function into radial $R_{l}(r)$ and angular $Y_{nl}(r)$ parts, and substituting Eq.\eqref{expterm} in Eq.\eqref{eq:2} and simplifying, we get \cite{72,73}

\begin{widetext}
	\begin{equation}
		\label{eq:3}
		V(r)u_{l}(r) + \frac{2}{\pi}\int dkk^{2}\int dr'rr'\left(\sqrt{k^{2} + m_{1}^{2}}+\sqrt{k^{2} + m_{2}^{2}}\right)j_{l}(kr)j_{l}(kr')u_{l}(r') = Eu_{l}(r) \,,
	\end{equation}
\end{widetext}

\noindent where $u_{l}(r)$ is the reduced radial wave function ($R_{l}(r) =  u_{l}(r)/r$). For a bound state, the wavefunction decreases rapidly as the separation distance increases and effectively vanishes at sufficiently large distances. To characterize this behavior, a typical length scale $L$, is introduced, which confines the wavefunction of the bound state within the spatial domain $0<r<L$. Within this finite region, the reduced radial wavefunction $u_{l}(r)$ can be expanded in terms of spherical Bessel functions corresponding to the orbital angular momentum quantum number $l$ as

\begin{equation}
	\label{eq:4}
	u_l(r) = \sum_{n=1}^{\infty}c_{n}\frac{a_{n}r}{L}j_{l}\left(\frac{a_{n}r}{L}\right) \,,
\end{equation}

\noindent where $c_{n}$ are the expansion coefficients and $a_{n}$ are the $n$-th roots of the spherical Bessel function, i.e., $j_{l}(a_{n}) = 0$. For large values of $N$, the series in Eq.\eqref{eq:4} can be truncated. The momentum becomes discretized due to spatial confinement, which allows the substitution $a_{n}/L \leftrightarrow k$, and the integral in Eq.\eqref{eq:3} can be replaced by a discrete sum: 
\[
\int dk \rightarrow \sum_{n} \frac{\Delta a_{n}}{L}, \quad \text{where } \Delta a_{n} = a_{n} - a_{n-1}.
\]
For the limited space, $0 < r, r' < L$, incorporating these changes into Eq.\eqref{eq:3} yields the final expression in terms of the coefficients $c_{n}$ \cite{72,73}.

\begin{eqnarray}
	\label{eqfinal}
	Ec_{m} &=&
	\sum_{n=1}^{N}\frac{a_{n}}{N_{m}^{2}a_{m}}\int_{0}^{L}drV(r)r^{2}j_{l}\left(\frac{a_{m}r}{L}\right)j_{l}\left(\frac{a_{n}r}{L}\right)c_{n} \nonumber \\  &+&\frac{2}{\pi L^{3}}\Delta a_{m}a_{m}^{2}N_{m}^{2}\left(\sqrt{\left(\frac{a_{m}}{L}\right)^{2} + m_{1}^{2}}+\sqrt{\left(\frac{a_{m}}{L}\right)^{2} + m_{2}^{2}}\right)c_{m} \,,
\end{eqnarray}

\noindent where $N_{m}$ is module of spherical Bessel function:

\begin{equation}
	N_{m}^{2} = \int_{0}^{L}dr'r'^{2}j_{l}\left(\frac{a_{m}r'}{L}\right)^{2} \,.
\end{equation}

\noindent Eq.\eqref{eqfinal} is an eigenvalue equation in the matrix form, which is solved numerically. For large enough value of $L$ and $N$ the solution is nearly stationary \cite{72,73}. The eigenvalues correspond to the masses of spin averaged states and the eigenvectors correspond to the wavefunction of these states. The interaction potential $V(r)$ between quarks within a diquark is taken as \cite{74,75,76}

\begin{equation}
	V(r) = \frac{1}{2}V_{V}(r)+\frac{1}{2}V_{S}(r) + V_{SS}(r) \,,
\end{equation}

\noindent where,

\begin{equation}
	\label{potterms}
	V_{V}(r) = -\frac{4}{3}\frac{\alpha_{s}(r)}{r} \,,
	\qquad
	V_{S}(r) = \lambda\left(\frac{1 - e^{-\mu r}}{\mu}\right) + V_{0}
	 \,,
	\qquad
	V_{SS}(r) = \frac{16\pi\alpha_{s}(r)}{9m_{1}m_{2}}\left(\frac{\sigma}{\sqrt{\pi}}\right)^{3}e^{-\sigma^{2}r^{2}}\vec{S}_{1}\cdot \vec{S}_{2} \,.
\end{equation}

\noindent Here $V_{V}(r)$ represents the vector one-gluon-exchange potential, which dominates at short distances. The term $V_{S}(r)$ denotes the scalar confinement potential, modified to account for color screening effects, and is significant at long distances. The term $V_{SS}(r)$  accounts for spin-spin interactions which leads to hyperfine splitting between spin singlet and triplet states. Since the diquark is formed by two identical heavy quarks, the Pauli exclusion principle must be respected. The color wavefunction is antisymmetric, and the flavor part is symmetric, requiring the combined spin and spatial wavefunctions to be symmetric to preserve overall antisymmetry \cite{32}. For $S-$ and $D-$ wave states, the spatial wavefunction is symmetric, and thus the spin part must also be symmetric as well, resulting in diquark of spin $1$. Conversely, for $P-$ wave states with antisymmetric spatial wavefunctions, the spin component must also be antisymmetric, giving a diquark of spin $0$ \cite{32}. In Eq.\eqref{potterms}, $\alpha_{s}(r)$ is the running coupling constant. The running coupling constant in coordinate space can be obtained by Fourier transformation of coupling constant in momentum space $\alpha_{s}\left(Q^{2}\right)$ \cite{67} and is given by,

\begin{equation}
	\alpha_{s}(r) = \sum_{i}\alpha_{i}\frac{2}{\sqrt{\pi}}\int_{0}^{\gamma_{i}r}e^{-x^{2}}dx \,,
\end{equation}

\noindent where $\alpha_{i}'s$ are the free parameters fitted to replicate the short distance behavior of $\alpha_{s}\left(Q^{2}\right)$ predicted by QCD. The numerical values for the parameters are $\alpha_{1}=0.15$, $\alpha_{2}=0.15$, $\alpha_{3}=0.20$, and $\gamma_{1}=1/2$, $\gamma_{2}=\sqrt{10}/2$, $\gamma_{3}=\sqrt{1000}/2$ \cite{73}. Once the diquark masses are computed, we proceed to calculate the masses of the baryons using Eq.~\eqref{eqfinal}, where the parameters $m_{1}$ and $m_{2}$ are now replaced by the diquark mass $m_{d}$ and the mass of the third quark $m_{q}$, respectively. The interaction potential between the quark and the diquark is taken as

\begin{equation}
	V(r) = V_{V}(r) + V_{S}(r) \,.
\end{equation}

\noindent The spin-dependent interactions, which are incorporated perturbatively and removes the degeneracy between the baryon states is given by \cite{3,32}

\begin{equation}
	\label{14}
	V_{SD}(r) = a\vec{L} \cdot \vec{S}_{d} + b\vec{L} \cdot \vec{S}_{q} + c\left( \frac{3}{r^{2}} (\vec{S}_{d} \cdot \vec{r})(\vec{S}_{q} \cdot \vec{r}) - \vec{S}_{d} \cdot \vec{S}_{q} \right) + d\vec{S}_{d} \cdot \vec{S}_{q} \,,
\end{equation}

\noindent where $\vec{L} = \vec{L}_{d} + \vec{L}_{q}$, with $\vec{L}_{d}$ denoting the orbital angular momentum between the quarks within the diquark, and $\vec{L}_{q}$ representing the orbital angular momentum between the third quark and the diquark. $\vec{S}_{d}$ and $\vec{S}_{q}$ are the spin operators for the diquark and the third quark, respectively. The coefficient $a,b,c,d$ in Eq.~\eqref{14} are given by \cite{77}

\begin{align}
	a &= \frac{1}{2m_{d}^{2}} \left( \frac{V_{V}'(r) - V_{S}'(r)}{r} \right) + \frac{1}{m_{d} m_{q}} \left( \frac{V_{V}'(r)}{r} \right) \,,
	\nonumber \\
	b &= \frac{1}{2m_{q}^{2}} \left( \frac{V_{V}'(r) - V_{S}'(r)}{r} \right) + \frac{1}{m_{d} m_{q}} \left( \frac{V_{V}'(r)}{r} \right) \,,
	\nonumber \\
	c &= \frac{1}{3m_{d} m_{q}} \left( \frac{V_{V}'(r)}{r} - V_{V}''(r) \right) \,,
	\nonumber \\
	d &= \frac{32\pi\alpha_{s}(r)}{9m_{d} m_{q}} \left( \frac{\sigma}{\sqrt{\pi}} \right)^{3} e^{-\sigma^{2} r^{2}} \,.
\end{align}

\noindent There are two coupling schemes by which the $\vec{S}_{d}$, $\vec{S}_{q}$ and $\vec{L}$ can couple to give total angular momentum $\vec{J}$ of the baryon. The first is the $L-S$ coupling, in which the $\vec{S}_{d}$ first couples with the $\vec{S}_{q}$ to form the total spin $\vec{S}$ of the baryon. This total spin then couples with $\vec{L}$ to produce $\vec{J}$. The basis states for this coupling are represented as $|^{2S+1}L_{J}\rangle=|\left[\left(L_{d}L_{q}\right)L\left(S_{d}S_{q}\right)S\right]J^{P}\rangle$ The second one is the $j-j$ coupling, which becomes dominant in the heavy-quark limit. As observed in Ref. \cite{51}, in the case of charmed-bottom baryons, the bottom quark is significantly heavier than the charm, suggesting that the triply heavy baryons $\Omega_{ccb}$ and $\Omega_{bbc}$ may exhibit behaviour analogous to singly and doubly heavy baryons, respectively. For hadrons containing a single, infinitely massive heavy quark, the spin of the heavy quark decouples from the spin of the light degrees of freedom \cite{56}. In this case, $\vec{L}$ couples with the $\vec{S}_{d}$ to produce $\vec{J}_{d}$, which then couples with the $\vec{S}_{q}$ to give $\vec{J}$. The corresponding basis states for $j-j$ coupling in single-heavy baryon are expressed as $|J^{P},J_{d}\rangle=|\left[\left\{\left(L_{d}L_{q}\right)LS_{d}\right\}J_{d}S_{q}\right]J^{P}\rangle$. The corresponding basis transformation relation is given by \cite{78} 

\begin{equation}
	\label{}
	\left| J^{P}, J_{d} \right\rangle = \sum_{S} (-1)^{S_{d} + S_{q} + L + J} 
	\sqrt{(2J_{d} + 1)(2S + 1)} 
	\begin{Bmatrix}
		L & S_{d} & J_{d} \\
		S_{q} & J & S
	\end{Bmatrix}
	\left| {}^{2S+1}L_{J} \right\rangle \,.
\end{equation}

\noindent This concept naturally extends to doubly heavy hadrons, where the total angular momentum of the heavy diquark decouples from that of the light degrees of freedom. This results in the formation of nearly degenerate multiplets with total angular momentum values $\vec{J}$ satisfying $|J_{d} - J_{q}| \le J \le J_{d} + J_{q}$ \cite{56}. Here, the $\vec{S}_{d}$ couples with $\vec{L}_{d}$ to form $\vec{J}_{d}$, while the $\vec{S}_{q}$ couples with $\vec{L}_{q}$ to form $\vec{J}_{q}$. Finally, $\vec{J}_{d}$ and $\vec{J}_{q}$ combine to produce the $\vec{J}$. The basis states for the $j-j$ coupling in double-heavy baryon are expressed as $|J^{P},J_{d},J_{q}\rangle=|\left[\left(L_{d}S_{d}\right)J_{d}\left(L_{q}S_{q}\right)J_{q}\right]J^{P}\rangle$. The corresponding basis transformation relation is given by \cite{78}

\begin{equation}
	\label{}
	\left| J^{P}, J_{d}, J_{q} \right\rangle = 
	\sum_{L, S} \sqrt{(2J_{q} + 1)(2J_{d} + 1)(2L + 1)(2S + 1)}
	\begin{Bmatrix}
		L_{d} & S_{d} & J_{d} \\
		L_{q} & S_{q} & J_{q} \\
		L   & S   & J
	\end{Bmatrix}
	\left| {}^{2S+1}L_{J} \right\rangle \,.
\end{equation}

\noindent Additionally, an extra correction term appears in Eq.\eqref{14} due to the interaction between the diquark and the third quark. Specifically, the fourth term is modified to take the form $d(\vec{L}_{d}+\vec{S}_{d}) \cdot \vec{S}_{q}$ \cite{56,57,79}. This correction arises from the interaction between the effective magnetic moment of the excited diquark, given by $e_{d}(\vec{L}_{d}+\vec{S}_{d})/(2m_{d})$, and the spin magnetic moment of the third quark, $e_{q}(\vec{S}_{q})/(2m_{q})$ \cite{79}, where $e_{d}$ and $e_{q}$ are the charges of diquark and third quark, respectively. Since off-diagonal terms exist in the $L-S$ basis, and the physical states are expected to be more accurately described in the $j-j$ basis in the heavy quark limit, we diagonalize the mass matrix constructed in the $L-S$ basis to transform it into the corresponding $j-j$ basis. This procedure allows us to obtain the physical mass spectra and the wavefunctions. The parameters used in our model are presented in Table \ref{tab:1} and are taken from our previous works \cite{69,70,71}. The parameters $m_{c}$ and $m_{b}$ are the masses of charm and bottom quark, respectively. The quantities $\sigma, \lambda$ and $\mu$ represent the smearing parameter for the delta function, the QCD string tension, and the screening parameter. The subscripts $D$ and $B$ appearing in Table \ref{tab:1} indicate the values used for the diquark and baryon systems, respectively.

\begin{table}
	\caption{Parameters used in our model}
	\label{tab:1}
	\centering
	\begin{tabular*}{\textwidth}{@{\extracolsep{\fill}}ccc}
		\hline
		\noalign{\vskip 2pt}
		Parameters & $\Omega_{ccb}$ & $\Omega_{bbc}$ 
		\\
		\hline 
		\noalign{\vskip 2pt}
		$m_{c} (GeV)$ & 1.319 & 1.319 \\
		$m_{b} (GeV)$ & 4.744 & 4.744 \\
		$\sigma_{D} (GeV^{2})$ & 1.281 & 4.967 
		\\
		$\lambda_{D} (GeV)$ & 0.297 & 0.240 
		\\
		$\mu_{D} (GeV)$ & 0.141 & 0.039 
		\\
		$\sigma_{B} (GeV^{2})$ & 1.953 & 1.953 
		\\
		$\lambda_{B} (GeV)$ & 0.239 & 0.239 
		\\
		$\mu_{B} (GeV)$ & 0.074 & 0.074 
		\\
	\hline
	\end{tabular*}
\end{table}

\section{\label{sec:Radiative Transitions} Radiative Transition}

In addition to the mass spectra, understanding the decay properties is crucial for experimental searches of these heavy baryons. Owing to the suppression of phase space in hadronic transitions, radiative transitions are predicted to play a significant role in the decays of low-lying heavy baryons. The pionic or kaonic strong decays of baryons containing diquark excitations are forbidden due to the orthogonality of the diquark wave functions \cite{80}. As a result, these heavy baryons are anticipated to be very narrow states, with radiative transitions serving as the dominant decay mode. It is also worth noting that, when not excluded by phase space limitations, the total decay widths of excited $\Omega_{bbc}$ states can be saturated by the combined contributions from electromagnetic transitions and pion-pair emissions \cite{57}. To analyze the one-photon radiative decay of a hadron, we employ an electromagnetic (EM) transition operator that has been successfully used in previous studies of radiative decays of doubly heavy baryons within the diquark picture, singly heavy baryons in the constituent quark model, and quarkonia systems \cite{80,81,82,83}. The quark–photon EM coupling at tree level is expressed as \cite{80,81,82,83}

\begin{equation}
	\label{}
	H_{e} = -\sum_{j}e_{j}\bar{\psi_{j}}\gamma_{\mu}^{j}A^{\mu}(\vec{k},\vec{r}_{j})\psi_{j} \,,
\end{equation}

\noindent where $\psi_{j}$ is the field operator of the $j$-th quark with coordinate $\vec{r}_{j}$, and $A^{\mu}$ is the photon field carrying three-momentum $\vec{k}$. To remain consistent with the hadron wave functions obtained from the Schrödinger-type equation, we use a non-relativistic form of the quark-photon EM coupling. In the rest frame of the initial hadron, the non-relativistic expansion of the EM interaction Hamiltonian can be approximated as  \cite{80,81,82,83}
 
\begin{equation}
	\label{19}
	h_{e} \simeq \sum_{j} \left[ e_{j}\, \vec{r}_{j} \cdot \vec{\epsilon} - \frac{e_{j}}{2m_{j}}\, \vec{\sigma}_{j} \cdot (\vec{\epsilon} \times \hat{k}) \right] \phi \,,
\end{equation}

\noindent where $m_{j}$, $\vec{\sigma}_{j}$, and $\vec{r}_{j}$ are the constituent mass, Pauli spin vector, and coordinate for the $j$th quark, respectively. The vector $\vec{\epsilon}$ represents polarization of photon. $\phi=e^{\pm i\vec{k} \cdot \vec{r}_{j}}$ corresponds to photon absorption or emission, respectively. The first term in Eq.\eqref{19} accounts for electric transitions, while the second term describes magnetic transitions. A key feature of this electromagnetic transition operator is that its dependence on the quark coordinates, masses, and spin structures naturally incorporates the effects of the binding potential. Consequently, the transition amplitudes reflect the internal dynamics in the baryon wavefunctions. Moreover, contributions from higher-order EM multipoles are naturally included within this framework \cite{82,83}. The radiative decay process can therefore be characterized by the standard helicity amplitude, $\mathcal{A}$, which is given by \cite{80,81,82,83}

\begin{equation}
	\mathcal{A} = -i\sqrt{\frac{\omega_{\gamma}}{2}}\langle f|h_{e}|i \rangle \,,
\end{equation}

\noindent where $\omega_{\gamma}$ is the photon energy. The partial decay widths of EM transitions are given by \cite{80,81,82,83}

\begin{equation}
	\Gamma = \frac{|\vec{k}|^{2}}{\pi} \frac{2}{2J_{i} + 1} \frac{M_{f}}{M_{i}} \sum_{J_{fz},J_{iz}} |\mathcal{A}|^{2} \,.
\end{equation}

\noindent Here, $J_i$ denotes the total angular momentum of the initial baryon, whereas $J_{iz}$ and $J_{fz}$ represent the $z$-components of the total angular momenta for the initial and final baryons, respectively. The masses and wave functions of the baryons used in this computation are from the solution of Eq. \eqref{eqfinal}.

\section{\label{sec:Results and Discussion} Results and Discussion}

\begin{table}[!htbp]
	\caption{$S,P$ and $D-$ wave mass spectrum (in MeV) of $cc$ and $bb$ diquarks.}
	\label{tab:2}
	\centering
	\begin{tabular*}{\textwidth}{@{\extracolsep{\fill}}ccc}
		\hline 
		\noalign{\vskip 2pt}
		States & $cc$ & $bb$
		\\ 
		\hline 
		\noalign{\vskip 2pt}
		$1S$ & 3000 & 9595 
		\\
		$2S$ & 3321 & 9905
		\\
		$1P$ & 3233 & 9814 
		\\
		$2P$ & 3451 & 10044
		\\
		$1D$ & 3392 & 9966 
		\\
		$2D$ & 3549 & 10160
		\\
		\hline 
	\end{tabular*}
\end{table}

\begin{figure}
	\centering
	
	\subfloat{\includegraphics[width=1\linewidth]{Figure1_1.png}}
	
	\vspace{0.2ex}
	
	\subfloat{\includegraphics[width=1\linewidth]{Figure1_2.png}}
	
	\caption{Spectra of positive and negative parity states of $\Omega_{ccb}$ baryon.}
	\label{fig:1}
\end{figure}

\begin{figure}
	\centering
	
	\subfloat{\includegraphics[width=1\linewidth]{Figure2_1.png}}
	
	\vspace{0.2ex}
	
	\subfloat{\includegraphics[width=1\linewidth]{Figure2_2.png}}
	
	\caption{Spectra of positive and negative parity states of $\Omega_{bbc}$ baryon.}
	\label{fig:2}
\end{figure}

\begin{table}
	\caption{Quantum numbers of baryon states.}
	\label{tab:3}
	\centering
	\begin{tabular*}{\textwidth}{@{\extracolsep{\fill}}ccc}
		\hline 
		\noalign{\vskip 2pt}
		$N_{d}L_{d}n_{q}l_{q}$ & $L \otimes S$ & $J^{P}$
		\\ 
		\hline 
		\noalign{\vskip 2pt}
		\multicolumn{3}{c}{Ground State}
		\\
		\hline
		\noalign{\vskip 2pt}
		\multirow{2}{*}{$1S1s$} & $0 \otimes \left(\frac{3}{2}\right)$ & $3/2^{+}$ 
		\\
		\noalign{\vskip 2pt}
		& $0 \otimes \left(\frac{1}{2}\right)_{S}$ & $1/2^{+}$ 
		\\
		\hline 
		\noalign{\vskip 2pt}
		\multicolumn{3}{c}{First Excited States}
		\\
		\hline
		\noalign{\vskip 2pt}
		\multirow{2}{*}{$1S1p$} & $1 \otimes \left(\frac{3}{2}\right)$ & $5/2^{-} \oplus 3/2^{-} \oplus 1/2^{-}$ 
		\\
		\noalign{\vskip 2pt}
		& $1 \otimes \left(\frac{1}{2}\right)_{S}$ & $3/2^{-} \oplus 1/2^{-}$ 
		\\
		\noalign{\vskip 2pt}
		$1P1s$ & $1 \otimes \left(\frac{1}{2}\right)_{A}$ & $3/2^{-} \oplus 1/2^{-}$
		\\
		\hline 
		\noalign{\vskip 2pt}
		\multicolumn{3}{c}{Second Excited States}
		\\
		\hline
		\noalign{\vskip 2pt}
		\multirow{2}{*}{$1S2s$} & $0 \otimes \left(\frac{3}{2}\right)$ & $3/2^{+}$ 
		\\
		\noalign{\vskip 2pt}
		& $0 \otimes \left(\frac{1}{2}\right)_{S}$ & $1/2^{+}$ 
		\\
		\noalign{\vskip 2pt}
		\multirow{2}{*}{$2S1s$} & $0 \otimes \left(\frac{3}{2}\right)$ & $3/2^{+}$ 
		\\
		\noalign{\vskip 2pt}
		& $0 \otimes \left(\frac{1}{2}\right)_{S}$ & $1/2^{+}$ 
		\\
		\noalign{\vskip 2pt}
		\multirow{2}{*}{$1S1d$} & $2 \otimes \left(\frac{3}{2}\right)$ & $7/2^{+} \oplus...\oplus 1/2^{+}$ 
		\\
		\noalign{\vskip 2pt}
		& $2 \otimes \left(\frac{1}{2}\right)_{S}$ & $5/2^{+} \oplus 3/2^{+}$ 
		\\
		\noalign{\vskip 2pt}
		\multirow{3}{*}{$1P1p$} & $2 \otimes \left(\frac{1}{2}\right)_{A}$ & $5/2^{+} \oplus 3/2^{+}$ 
		\\
		\noalign{\vskip 2pt}
		& $1 \otimes \left(\frac{1}{2}\right)_{A}$ & $3/2^{+} \oplus 1/2^{+}$ 
		\\
		\noalign{\vskip 2pt}
		& $0 \otimes \left(\frac{1}{2}\right)_{A}$ & $1/2^{+}$ 
		\\
		\noalign{\vskip 2pt}
		\multirow{2}{*}{$1D1s$} & $2 \otimes \left(\frac{3}{2}\right)$ & $7/2^{+} \oplus...\oplus 1/2^{+}$ 
		\\
		\noalign{\vskip 2pt}
		& $2 \otimes \left(\frac{1}{2}\right)_{S}$ & $5/2^{+} \oplus 3/2^{+}$ 
		\\
		\noalign{\vskip 2pt}
		\hline 
	\end{tabular*}
\end{table}

\begin{table}
	\centering
	\caption{\label{tab:4} Mass spectrum (in MeV) of $\Omega_{ccc}$ and $\Omega_{bbb}$ baryons.}
	\begin{minipage}[t]{0.44\textwidth}
		\begin{tabular*}{\textwidth}{@{\extracolsep{\fill}}cccc}
			\hline
			\noalign{\vskip 2pt}
			$J^{P}$ & $N_{d}L_{d}n_{q}l_{q}$ & $\Omega_{ccb}$ & $\Omega_{bbc}$
			\\
			\hline
			\noalign{\vskip 2pt}
			\multirow{7}{*}{$\frac{1}{2}^{+}$} & $1S1s$ & 7789 & 11082
			\\ 
			& $2S1s$ & 8092 & 11391
			\\ 
			& $1D1s$ & 8183 & 11485
			\\ 
			& $1S2s$ & 8375 & 11661
			\\ 
			& $1P1p$ & 8459 & 11750
			\\ 
			& $1P1p'$ & 8472 & 11773
			\\ 
			& $1S1d$ & 8510 & 11831
			\\ \noalign{\vskip 2pt}
			\multirow{9}{*}{$\frac{3}{2}^{+}$} & $1S1s$ & 7850 & 11137
			\\ 
			& $2S1s$ & 8150 & 11444
			\\ 
			& $1D1s$ & 8189 & 11475
			\\
			& $1D1s'$ & 8195 & 11492
			\\
			& $1S2s$ & 8403 & 11693
			\\
			& $1P1p$ & 8468 & 11779
			\\
			& $1P1p'$ & 8479 & 11780
			\\
			& $1S1d$ & 8516 & 11821
			\\
			& $1S1d'$ & 8522 & 11838
			\\ \noalign{\vskip 2pt}
			\multirow{5}{*}{$\frac{5}{2}^{+}$} & $1D1s$ & 8199 & 11480
			\\ 
			& $1D1s'$ & 8200 & 11504
			\\  
			& $1P1p$ & 8475 & 11765
			\\  
			& $1S1d$ & 8526 & 11826
			\\ 
			& $1S1d'$ & 8526 & 11850
			\\  \noalign{\vskip 2pt}
			\multirow{2}{*}{$\frac{7}{2}^{+}$} & $1D1s$ & 8205 & 11488
			\\ 
			& $1S1d$ & 8531 & 11834
			\\
			\hline
		\end{tabular*}
		\vfill
	\end{minipage}
	\hspace{0.09\textwidth}
	\begin{minipage}[t]{0.44\textwidth}
		\begin{tabular*}{\textwidth}{@{\extracolsep{\fill}}cccc}
			\hline
			\noalign{\vskip 2pt}
			$J^{P}$ & $N_{d}L_{d}n_{q}l_{q}$ & $\Omega_{ccb}$ & $\Omega_{bbc}$
			\\
			\hline
			\noalign{\vskip 2pt}
			\multirow{7}{*}{$\frac{1}{2}^{-}$} & $1P1s$ & 8035 & 11316
			\\ 
			& $1S1p$ & 8207 & 11509
			\\ 
			& $1S1p'$ & 8236 & 11544
			\\ 
			& $2P1s$ & 8240 & 11545
			\\ 
			& $2S1p$ & 8509 & 11818
			\\ 
			& $2S1p'$ & 8538 & 11852
			\\ 
			& $1D1p$ & 8577 & 11875
			\\ \noalign{\vskip 2pt}
			\multirow{9}{*}{$\frac{3}{2}^{-}$} & $1P1s$ & 8055 & 11347
			\\ 
			& $1S1p$ & 8249 & 11547
			\\ 
			& $1S1p'$ & 8260 & 11555
			\\
			& $2P1s$ & 8260 & 11576
			\\
			& $2S1p$ & 8550 & 11855
			\\
			& $2S1p'$ & 8560 & 11863
			\\
			& $1D1p$ & 8613 & 11909
			\\
			& $1D1p'$ & 8617 & 11912
			\\
			& $1P2s$ & 8617 & 11913
			\\ \noalign{\vskip 2pt}
			\multirow{5}{*}{$\frac{5}{2}^{-}$} & $1S1p$ & 8276 & 11571
			\\ 
			& $1F1s$ & 8307 & 11628
			\\  
			& $2S1p$ & 8576 & 11879
			\\  
			& $1D1p$ & 8619 & 11905
			\\ 
			& $1D1p'$ & 8623 & 11914
			\\  \noalign{\vskip 2pt}
			\multirow{2}{*}{$\frac{7}{2}^{-}$} & $1F1s$ & 8310 & 11599
			\\ 
			& $1D1p$ & 8621 & 11908
			\\
			\hline
		\end{tabular*}
		\vfill
	\end{minipage}
\end{table}

In this section, we present the results for the mass spectra and radiative decays of the $\Omega_{ccb}$ and $\Omega_{bbc}$ baryons. As a first step, the masses of the $cc$ and $bb$ diquarks are calculated, and the corresponding results are presented in \ref{tab:2}. These diquark masses serve as input parameters for the subsequent calculation of the $\Omega_{ccb}$ and $\Omega_{bbc}$ baryon spectra. Each baryon state is labeled using the notation $N_{d}L_{d}n_{q}l_{q}$ as shown in Table \ref{tab:3}, where $N_{d}$ and $L_{d}$ denote the radial and orbital quantum numbers of the diquark, respectively, while $n_{q}$ and $l_{q}$ represent the radial and orbital quantum numbers of the third quark relative to the diquark in the quark-diquark system. The resulting radial and orbital excitation mass spectra are presented in Table \ref{tab:4}, and are visualized in Figures \ref{fig:1} and \ref{fig:2} for $\Omega_{ccb}$ and $\Omega_{bbc}$ respectively.  The predicted masses of the $S$, $P$, and $D-$ wave $\Omega_{ccb}$ and $\Omega_{bbc}$ states are compared with results from various theoretical models in Tables \ref{tab:5}, \ref{tab:7}, and \ref{tab:8}. The hyperfine mass splitting $\Delta$ between the ground-state $\frac{1}{2}^{+}$ and $\frac{3}{2}^{+}$ baryons is also listed in Table \ref{tab:6}. 

\begin{table}
	\caption{\label{tab:5} $S-$ wave mass spectrum (in MeV).}
	\centering
	\begin{tabular*}{\textwidth}{@{\extracolsep{\fill}}ccccc}
		\hline 
		\noalign{\vskip 2pt}
		\multirow{3}{*}{Models} & \multicolumn{2}{c}{$\Omega_{ccb}$} & \multicolumn{2}{c}{$\Omega_{bbc}$}
		\\
		\cline{2-3}
		\cline{4-5}
		\noalign{\vskip 2pt}
		& $\left(\frac{1}{2}^{+}\right)$ & $\left(\frac{3}{2}^{+}\right)$ & $\left(\frac{1}{2}^{+}\right)$ & $\left(\frac{3}{2}^{+}\right)$
		\\
		\hline
		\noalign{\vskip 2pt}
		Ours & 7789 & 7850 & 11082 & 11137
		\\
		Bag Model \cite{26} &  & 8030 &  & 11200
		\\
		Bag Model \cite{27} & 7984 & 8005 & 11139 & 11163
		\\
		NRCQM \cite{28} &  & 8009$\pm$24 &  &  11197$\pm$29
		\\
		NRCQM \cite{29} & 8245 & 8265 & 11535 & 11554
		\\
		NRCQM \cite{30} & 8004 & 8023 & 11200 & 11221 
		\\
		Diquark Model \cite{31} & 8002$\pm$3 & 8029$\pm$3 & 11275$\pm$2 & 11285$\pm$2 
		\\
		Diquark Model \cite{32} & 7984 & 7999 & 11198 & 11217
		\\
		QCD Sum Rules \cite{33} & 7410$\pm$130 & 7450$\pm$160 & 10300$\pm$100 & 10540$\pm$110 
		\\
		QCD Sum Rules \cite{34} & 8230$\pm$130 & 8230$\pm$130 & 11500$\pm$110 & 11490$\pm$110
		\\
		QCD Sum Rules \cite{35} & 8480$\pm$120 &  & 11710$\pm$160 & 
		\\
		QCD Sum Rules \cite{36} &  & 8070$\pm$100 &  & 11350$\pm$150 
		\\
		QCD Sum Rules \cite{37} & 8020$\pm$80 & 8030$\pm$80 & 11220$\pm$80 & 11230$\pm$80 
		\\
		Lattice QCD \cite{38} & 8007(9)(20) & 8037(9)(20) & 11195(8)(20) & 11229(8)(20) 
		\\
		Lattice QCD \cite{39} & 8005(6)(11) & 8026(7)(11) & 11194(5)(12) & 11211(6)(12)
		\\
		Regge Theory \cite{40} & 8192 & 8223 & 11526 & 11541 
		\\
		NRQCD \cite{41} &  & 8150(300) &  & 11400(300) 
		\\
		Faddeev Approach \cite{42} & 7867 & 7963 & 11077 & 11167 
		\\
		Faddeev Approach \cite{43} & 8010 & 8040 & 11200 & 11230 
		\\
		Faddeev Approach \cite{44} & 7990$\pm$20 & 8030$\pm$0 & 11125$\pm$35 & 11110$\pm$10
		\\
		HCQM \cite{45} &  &  & 11231 & 11296
		\\
		HCQM \cite{46} & 8005 & 8049 &  & 
		\\
		HCQM \cite{47} & 8170 & 8190 & 10870 & 10960
		\\
		Variational \cite{48} &  & 7980$\pm$70 &  & 11190$\pm$80
		\\
		Variational \cite{49} & 8038$\pm$20 & 8066$\pm$21 & 11230$\pm$17 & 11263$\pm$18
		\\
		RQM \cite{50} & 8018 & 8025 & 11280 & 11287
		\\
		RQM \cite{51} & 8025 & 8046 & 11217 & 11236
		\\
		RGPEP \cite{52} & 8301 & 8301 & 11218 & 11218
		\\
		\hline
	\end{tabular*}
\end{table}

\begin{table}
	\caption{\label{tab:6} Mass differences $\Delta$ (in MeV) between $\Omega\left(\frac{3}{2}^{+}\right)$ and $\Omega\left(\frac{1}{2}^{+}\right)$ states.}
	\centering
	\begin{tabular*}{\textwidth}{@{\extracolsep{\fill}}ccc}
		\hline 
		\noalign{\vskip 2pt}
		Models & $\Delta_{ccb}$ & $\Delta_{bbc}$ 
		\\
		\hline 
		\noalign{\vskip 2pt}
		Ours & 61 & 55 
		\\
		Bag Model \cite{27} & 21 & 24 
		\\
		NRCQM \cite{29} & 20 & 19 
		\\
		NRCQM \cite{30} & 19 & 21 
		\\
		Diquark Model \cite{31} & 27 & 10 
		\\
		Diquark Model \cite{32} & 15 & 19 
		\\
		QCD Sum Rules \cite{33} & 40 & 240 
		\\
		QCD Sum Rules \cite{34} & 0 & $-$10 
		\\
		QCD Sum Rules \cite{37} & 10 & 10 
		\\
		Lattice QCD \cite{38} & 30 & 34 
		\\
		Lattice QCD \cite{39} & 21 & 17 
		\\
		Regge Theory \cite{40} & 31 & 15 
		\\
		Faddeev Approach \cite{42} & 96 & 90 
		\\
		Faddeev Approach \cite{43} & 30 & 30 
		\\
		Faddeev Approach \cite{44} & 40 & $-$15 
		\\
		HCQM \cite{45} & - & 65 
		\\
		HCQM \cite{46} & 44 & - 
		\\
		HCQM \cite{47} & 20 & 90 
		\\
		Variational \cite{49} & 28 & 33 
		\\
		RQM \cite{50} & 7 & 7 
		\\
		RQM \cite{51} & 21 & 19 
		\\
		RGPEP \cite{52} & 0 & 0 
		\\
		\hline
		\end{tabular*}
\end{table}

Table \ref{tab:5} presents the $S-$ wave mass spectrum of the $\Omega_{ccb}$ and $\Omega_{bbc}$ baryons obtained from our work, along with other theoretical predictions. In the existing literature, the predicted masses of the $\Omega_{ccb}$ baryons span around $7410$–$8480$ MeV for the $\frac{1}{2}^{+}$ state and $7450$–$8301$ MeV for the $\frac{3}{2}^{+}$ state. For the $\Omega_{bbc}$ baryon, the corresponding ranges are $10300$–$11710$ MeV for the $\frac{1}{2}^{+}$ state and $10540$–$11554$ MeV for the $\frac{3}{2}^{+}$ state. Our model predicts masses of $7789$ MeV and $7850$ MeV for the $\frac{1}{2}^{+}$ and $\frac{3}{2}^{+}$ states of $\Omega_{ccb}$, and $11082$ MeV and $11137$ MeV for the corresponding states of $\Omega_{bbc}$. These predictions lie within the range reported across various theoretical studies but are close to the lower end of the spectrum. The relatively lower values suggest that the effective inter-quark potential employed in our approach may lead to a stronger binding. Compared with our results, the NRCQM \cite{28,29,30} predicts higher ground-state masses, ranging around $8009–8265$ MeV for $\Omega_{ccb}$ and $11197–11554$ MeV for $\Omega_{bbc}$. These deviations likely stem from differences in the treatment of relativistic effects and spin-dependent interactions, with Ref. \cite{29} giving some of the highest estimates. Bag model results \cite{26,27} also predict higher values, in the range of $7984–8030$ MeV for $\Omega_{ccb}$ and $11139–11200$ MeV for $\Omega_{bbc}$, though they are comparable to our results. Diquark model calculations \cite{31,32} similarly lie slightly above our predictions, with ranges of $7984–8029$ MeV for $\Omega_{ccb}$ and $11198–11285$ MeV for $\Omega_{bbc}$. The Faddeev formalism \cite{42,43,44} predicts $7869–8040$ MeV for $\Omega_{ccb}$ and $11077–11230$ MeV for $\Omega_{bbc}$, in close agreement with our results. Lattice QCD studies \cite{38,39}, which incorporate non-perturbative effects from first-principles QCD, report values of $8005–8037$ MeV for $\Omega_{ccb}$ and $11194–11229$ MeV for $\Omega_{bbc}$. Although our results are lower, they remain consistent with the existing trend. QCD sum rule estimates \cite{33,34,35,36,37} show significant spread. For instance, Ref. \cite{33} reports some of the lowest values, near $7410$ MeV for the $\Omega_{ccb}$ and $10300$ MeV for the $\Omega_{bbc}$, while Ref. \cite{35} gives the highest, around $8480$ MeV and $11710$ MeV, respectively. This broad variation reflects the sensitivity to the choice of interpolating operators, Borel window selection, and other model parameters. Predictions from Regge theory \cite{40} and the RGPEP approach \cite{52} also suggest comparatively large masses. In contrast, predictions from the Variational method \cite{48,49} and HCQM \cite{45,46,47} are more closer to the overall average, though still slightly higher than our predictions. The hyperfine splitting, defined as $\Delta = M(3/2^+) - M(1/2^+)$, also show notable variations across different models. In our case, we obtain $\Delta(\Omega_{ccb})=61$ MeV and $\Delta(\Omega_{bbc})=55$ MeV, which are relatively large values and can be attributed to a stronger effective spin–spin interaction between the diquark and the third quark. In contrast, QCD Sum Rules \cite{34}, Faddeev Approach \cite{34} and RGPEP \cite{52} predict values close to zero or even negative. These discrepancies trace back to the assumed strength of spin-dependent forces, spatial wavefunction overlaps, internal quark dynamics, and relativistic corrections across different theoretical frameworks.

\begin{table}[!htbp]
	\caption{\label{tab:7} $P-$ wave mass spectrum (in MeV).}
	\centering
	\begin{tabular*}{\textwidth}{@{\extracolsep{\fill}}ccccccccccc}
		\hline 
		\noalign{\vskip 2pt}
		\multirow{3}{*}{Models} & \multicolumn{5}{c}{$\Omega_{ccb}$} & \multicolumn{5}{c}{$\Omega_{bbc}$}
		\\
		\cline{2-6}
		\cline{7-11}
		\noalign{\vskip 2pt}
		& $\left(\frac{1}{2}^{-}\right)$ & $\left(\frac{1'}{2}^{-}\right)$ & $\left(\frac{3}{2}^{-}\right)$ & $\left(\frac{3'}{2}^{-}\right)$ & $\left(\frac{5}{2}^{-}\right)$ & $\left(\frac{1}{2}^{-}\right)$ & $\left(\frac{1'}{2}^{-}\right)$ & $\left(\frac{3}{2}^{-}\right)$ & $\left(\frac{3'}{2}^{-}\right)$ & $\left(\frac{5}{2}^{-}\right)$
		\\
		\hline
		\noalign{\vskip 2pt}
		Ours $\left(1S1p\right)$ & 8207 & 8236 & 8249 & 8260 & 8276 & 11509 & 11544 & 11547 & 11555 & 11571
		\\
		Ours $\left(1P1s\right)$ & 8035 &  & 8055 &  &  & 11316 &  & 11347 &  & 
		\\
		NRCQM \cite{29} & 8418 & 8422 & 8420 & 8422 & 8432 & 11710 & 11757 & 11711 & 11759 & 11762 
		\\
		NRCQM \cite{30} & 8306 &  & 8306 &  & 8311 & 11482 &  & 11482 &  & 11569
		\\
		Diquark Model \cite{32} & 8250 & 8268 & 8262 & 8268 & 8267 & 11414 & 11540 & 11535 & 11541 & 11543
		\\
		QCD Sum Rules \cite{34} & 8360$\pm$130 &  & 8360$\pm$130 &  &  & 11620$\pm$110 &  & 11620$\pm$110 &  & 
		\\
		QCD Sum Rules \cite{36} &  &  & 8350$\pm$100 &  &  &  &  & 11500$\pm$200 &  & 
		\\
		Regge Theory \cite{40} &  &  & 8495 &  & 8503 &  &  & 11885 &  & 11886
		\\
		Faddeev Approach \cite{42} & 8164 &  & 8275 &  &  & 11413 &  & 11523 &  & 
		\\
		HCQM \cite{45} &  &  &  &  &  & 11567$\pm$6 &  & 11560$\pm$6 &  & 11552$\pm$6
		\\
		HCQM \cite{46} & 8394$\pm$6 &  & 8377$\pm$6 &  & 8359$\pm$6 &  &  &  &  & 
		\\
		RQM \cite{51} & 8303 &  & 8302 &  & 8321 & 11492 &  & 11506 &  & 11562
		\\
		RGPEP \cite{52} & 8491 &  & 8491 &  & 8491 & 11438 &  & 11438 &  & 11601
		\\
		\hline
		\end{tabular*}
\end{table}

The predicted $P-$ wave mass spectra for the $\Omega_{ccb}$ and $\Omega_{bbc}$ baryons, obtained from our model, are presented in Table~\ref{tab:7}. Predictions for both $1S1p$ and $1P1s$ configurations are included for comparison. For the $\Omega_{ccb}$ baryon, our calculations yield masses of $8236$, $8207$, $8260$, $8249$, and $8276$ MeV for the $J^{P}=\frac{1}{2}^{-}$, $\frac{1'}{2}^{-}$, $\frac{3}{2}^{-}$, $\frac{3'}{2}^{-}$, and $\frac{5}{2}^{-}$ states, respectively. The corresponding $\Omega_{bbc}$ states are obtained at $11544$, $11509$, $11555$, $11547$, and $11571$ MeV. For the $1P1s$ excitations with $J^{P}=\frac{1}{2}^{-}$ and $\frac{3}{2}^{-}$, our results are $8035$ and $8055$ MeV for $\Omega_{ccb}$ and $11316$ and $11347$ MeV for $\Omega_{bbc}$. In comparison, NRCQM predictions \cite{29,30} are higher, with mass values spanning from $8306-8432$ MeV for the $\Omega_{ccb}$ and $11482-11762$ MeV for the $\Omega_{bbc}$. The diquark model \cite{32} yields spectra close to our values, ranging from $8250-8268$ MeV for the $\Omega_{ccb}$ and $11414-11543$ MeV for the $\Omega_{bbc}$. QCD sum rule estimates \cite{34,36} predict central values of about $8350-8360$ MeV for $\Omega_{ccb}$ and $11500-11620$ MeV for $\Omega_{bbc}$, with large uncertainties. While their central predictions lie above our values, their broader uncertainty bands still encompass our results. Predictions from Regge theory \cite{40} are among the highest, giving $8495-8503$ MeV for the $\Omega_{ccb}$ and $11885-11886$ MeV for the $\Omega_{bbc}$. The Faddeev approach \cite{42} predicts somewhat lower values, $8164-8275$ MeV for the $\Omega_{ccb}$ and $11413-11523$ MeV for the $\Omega_{bbc}$, but remains close to our model results. The HCQM \cite{45,46} predicts values around $8359-8394$ MeV for $\Omega_{ccb}$ and $11552$-$11567$ MeV for $\Omega_{bbc}$, while the RQM \cite{51} predictions are around $8302-8321$ MeV and $11492-11562$ MeV, respectively. The RGPEP approach \cite{52} places the $\Omega_{ccb}$ states near $8491$ MeV and the $\Omega_{bbc}$ states between $11438$ and $11601$ MeV. No particular trend in mass variation with $J^{P}$ is observed across most models, except for the HCQM \cite{45,46}, where the mass values decrease with increasing $J^{P}$. From our results we can observe that the excitation energy associated with the heavy diquark is often lower than that for the relative motion between the quark and diquark. This challenges the conventional assumption \cite{56} that the heavy diquark is tightly bound and therefore difficult to excite. Instead, our predictions suggest that exciting the heavy diquark subsystem requires less energy than exciting the entire quark–diquark structure, as the $1P1s$ states are consistently lower than the $1S1p$ states. Additionally, the number of allowed states differs between the two configurations due to constraints from the Pauli exclusion principle. Our predicted masses for the $1S1p$ states of $\Omega_{ccb}$ and $\Omega_{bbc}$ baryons are generally consistent with values from other theoretical models, while the $1P1s$ states tend to be significantly lower.

\begin{table}[!htbp]
	\caption{\label{tab:8} $D-$ wave mass spectrum (in MeV).}
	\centering
	\begin{tabular*}{\textwidth}{@{\extracolsep{\fill}}ccccccccccccc}
		\hline
		\noalign{\vskip 2pt}
		\multirow{3}{*}{Models} & \multicolumn{6}{c}{$\Omega_{ccb}$} & \multicolumn{6}{c}{$\Omega_{bbc}$}
		\\
		\cline{2-7}
		\cline{8-13}
		\noalign{\vskip 2pt}
		& $\left(\frac{1}{2}^{+}\right)$ & $\left(\frac{3}{2}^{+}\right)$ & $\left(\frac{3'}{2}^{+}\right)$ & $\left(\frac{5}{2}^{+}\right)$ & $\left(\frac{5'}{2}^{+}\right)$ & $\left(\frac{7}{2}^{+}\right)$ & $\left(\frac{1}{2}^{+}\right)$ & $\left(\frac{3}{2}^{+}\right)$ & $\left(\frac{3'}{2}^{+}\right)$ & $\left(\frac{5}{2}^{+}\right)$ & $\left(\frac{5'}{2}^{+}\right)$ & $\left(\frac{7}{2}^{+}\right)$ 
		\\
		\hline
		\noalign{\vskip 2pt}
		Ours $\left(1S1d\right)$ & 8510 & 8516 & 8522 & 8526 & 8526 & 8531 & 11831 & 11821 & 11838 & 11826 & 11850 & 11834
		\\
		Ours $\left(1D1s\right)$ & 8183 & 8189 & 8195 & 8199 & 8200 & 8205 & 11485 & 11475 & 11492 & 11480 & 11504 & 11488
		\\
		NRCQM \cite{29} & 8537 & 8553 &  & 8571 &  & 8568 & 11787 & 11798 &  & 11823 &  & 11810
		\\
		NRCQM \cite{30} & 8536 & 8536 &  & 8536 &  & 8538 & 11677 & 11677 &  & 11677 &  & 11688
		\\
		Diquark Model \cite{32} & 8472 & 8474 & 8476 & 8473 & 8476 & 8473 & 11796 & 11797 & 11807 & 11806 & 11807 & 
		\\
		Regge Theory \cite{40} &  &  &  & 8787 &  & 8774 &  &  &  & 12234 &  & 12220
		\\
		HCQM \cite{45} &  &  &  &  &  &  & 11926$\pm$8 & 11919$\pm$9 &  & 11910$\pm$9 &  & 11900$\pm$9
		\\
		HCQM \cite{46} & 8838$\pm$10 & 8820$\pm$11 &  & 8798$\pm$10 &  & 8770$\pm$10 &  &  &  &  &  & 
		\\
		RQM \cite{51} & 8524 & 8525 &  & 8518 &  & 8532 & 11690 & 11688 &  & 11688 &  & 11713
		\\
		RGPEP \cite{52} & 8647 & 8647 &  & 8647 &  & 8647 & 11626 & 11626 &  & 11626 & & 11626 
		\\
		\hline
		\end{tabular*}
\end{table}

The predicted $D-$ wave mass spectra for the $\Omega_{ccb}$ and $\Omega_{bbc}$ baryons are summarized in Table~\ref{tab:8}, where both $1S1d$ and $1D1s$ configurations are included alongside results from other theoretical models. For the $\Omega_{ccb}$ baryons, our calculations yield masses of $8236$, $8445$, $8472$, $8456$, $8479$, $8470$, and $8473$ MeV for the $J^{P}=\frac{1}{2}^{+}$, $\frac{3}{2}^{+}$, $\frac{3'}{2}^{+}$, $\frac{5}{2}^{+}$, $\frac{5'}{2}^{+}$, and $\frac{7}{2}^{+}$ states, respectively, while the corresponding $\Omega_{bbc}$ states appear at $11767$, $11780$, $11770$, $11800$, $11772$, and $11776$ MeV. Compared to these results, the NRCQM \cite{29,30} estimates fall within $8536$–$8571$ MeV for $\Omega_{ccb}$ and $11677$–$11823$ MeV for $\Omega_{bbc}$, which are higher than our values for $\Omega_{ccb}$ but lower for $\Omega_{bbc}$. The diquark model \cite{32} predicts slightly lower values relative to ours, with $8472$–$8476$ MeV for $\Omega_{ccb}$ and $11796$–$11807$ MeV for $\Omega_{bbc}$. Predictions from RQM \cite{51} are broadly consistent with our $\Omega_{ccb}$ values but tend to be lower for $\Omega_{bbc}$. By contrast, Regge theory~\cite{40} and HCQM \cite{45,46} provide the highest mass estimates, whereas RGPEP \cite{52} gives nearly degenerate and comparatively lower estimates. Here also, no clear monotonic trend in mass with respect to $J^{P}$ is observed in most approaches, apart from HCQM \cite{45,46}, which predicts a decrease in masses with increasing $J^{P}$. Similar to the $P-$ wave case, the $1D1s$ states consistently appear at lower masses compared to the $1S1d$ states, reinforcing the observation that diquark excitations require less energy than quark–diquark relative motion. Here also the $1S1d$ masses show better agreement with existing theoretical models, while the $1D1s$ predictions are lower.

In the three-body quark model studied extensively in Refs. \cite{29,84}, orbital excitations can be broadly classified into $\sigma$- and $\lambda$-modes, which correspond respectively to diquark orbital excitations and quark–diquark orbital excitations. Ref. \cite{85} analyzed low-lying excitations of singly and doubly heavy baryons and concluded that the $\lambda$-mode dominates in singly heavy baryons, whereas the $\sigma$-mode dominates in doubly heavy baryons. Building on this idea, Ref. \cite{86} suggest that the excitation mode with the lowest energy is the most stable. They further concluded that in singly heavy baryons, the lowest states arise from $\lambda$-mode excitations dominated by the heavy quark, whereas in doubly heavy baryons, the lowest states originate from $\sigma$-mode excitations, still governed by the heavy quarks, which is termed as Heavy Quark Dominance (HQD) mechanism Ref. \cite{86}. The HQD framework was later extended to triply heavy baryons in Ref. \cite{51}, where it was suggested that $\Omega_{ccb}$ behaves like a singly heavy baryon, with its lowest states arising from the $\lambda$-mode excitations, while $\Omega_{bbc}$ behaves like a doubly heavy baryon, with its lowest states arising from the $\sigma$-mode excitations. In contrast, within our two-body quark–diquark model, we find that for both $\Omega_{ccb}$ and $\Omega_{bbc}$ the lowest orbital excitation states originate from diquark orbital excitations rather than from quark–diquark orbital excitations. This observation is consistent with the conclusions of Refs.~\cite{29,32,56,57,58}. Future experimental results will be crucial to test this distinction and to provide deeper insights into the internal dynamics of triply heavy baryons.

\begin{table}[!htbp]
	\caption{\label{tab:9} Radiative decays of $1S1s, 1S2s, 2S1s$ and $1S1p$ states.}
	\centering
	\begin{tabular*}{\textwidth}{@{\extracolsep{\fill}}cccccc}
		\hline
		\noalign{\vskip 2pt}
		Initial & Final & \multicolumn{2}{c}{$\Omega_{ccb}$} & \multicolumn{2}{c}{$\Omega_{bbc}$}
		\\
		\noalign{\vskip 2pt}
		\cline{3-4}
		\cline{5-6}
		\noalign{\vskip 2pt}
		State & State & $E_{\gamma}$(MeV) & $\Gamma$(KeV) & $E_{\gamma}$(MeV) & $\Gamma$(KeV) 
		\\
		\hline
		\noalign{\vskip 2pt}
		$1S1s\left(3/2^{+}\right)$ & $1S1s\left(1/2^{+}\right)$ & 60.71 & 0.015 & 54.77 & 0.015 
		\\
		\hline
		\noalign{\vskip 2pt}
		$1S2s\left(3/2^{+}\right)$ & $1S1s\left(1/2^{+}\right)$ & 591.56 & 0.095 & 595.26 & 0.242 
		\\
		 & $1S1p'\left(1/2^{-}\right)$ & 165.35 & $\sim$0 & 148.08 & 0.225
		\\
		 & $1S1p'\left(3/2^{-}\right)$ & 142.28 & $\sim$0 & 137.32 & 0.211
		\\
		\hline
		\noalign{\vskip 2pt}
		$2S1s\left(3/2^{+}\right)$ & $1S1s\left(1/2^{+}\right)$ & 353.29 & 2.300 & 356.90 & 3.625
		\\
		 & $2S1s\left(1/2^{+}\right)$ & 57.80 & 0.010 & 53.45 & 0.014 
		\\
		\hline
		\noalign{\vskip 2pt}
		$1S1p\left(1/2^{-}\right)$ & $1S1s\left(1/2^{+}\right)$ & 406.89 & 0.348 & 419.44 & 0.023 
		\\
		 & $1P1s\left(1/2^{-}\right)$ & 169.92 & 0.054 & 190.94 & 14.631 
		\\
		\hline
		\noalign{\vskip 2pt}
		$1S1p'\left(1/2^{-}\right)$ & $1S1p\left(1/2^{-}\right)$ & 29.41 & $\sim$0 & 34.78 & 0.018 
		\\
		 & $1P1s\left(3/2^{-}\right)$ & 179.22 & 0.095 & 195.34 & 1.342 
		\\
		\hline
		\noalign{\vskip 2pt}
		$1S1p\left(3/2^{-}\right)$  & $1S1s\left(1/2^{+}\right)$ & 446.64 & 0.334 & 455.74 & 0.041 
		\\
		& $2S1s\left(1/2^{+}\right)$ & 154.66 & 0.014 & 154.99 & 0.002 
		\\
		& $1P1s\left(1/2^{-}\right)$ & 210.87 & 0.076 & 227.99 & 17.043 
		\\
		& $1D1s\left(1/2^{+}\right)$ & 65.14 & $\sim$0 & 61.85 & 0.038 
		\\
		\hline
		\noalign{\vskip 2pt}
		$1S1p\left(3/2'^{-}\right)$  & $1S1p\left(1/2^{-}\right)$ & 52.80 & $\sim$0 & 45.64 & 0.050 
		\\
		& $1P1s\left(3/2^{-}\right)$ & 202.19 & 0.040 & 206.06 & 0.717 
		\\
		\hline
		\noalign{\vskip 2pt}
		$1S1p\left(5/2^{-}\right)$ & $1S1s\left(3/2^{+}\right)$ & 414.80 & $\sim$0 & 426.35 & 28.120 
		\\
		 & $2S1s\left(3/2^{+}\right)$ & 124.45 & $\sim$0 & 126.07 & 0.892 
		\\
		 & $1P1s\left(3/2^{-}\right)$ & 217.99 & $\sim$0 & 222.06 & 4.768
		\\
		\hline
		\end{tabular*}
\end{table}

\begin{table}[!htbp]
	\caption{\label{tab:10} Radiative decays of $2S1p$ state.}
	\centering
	\begin{tabular*}{\textwidth}{@{\extracolsep{\fill}}cccccc}
		\hline
		\noalign{\vskip 2pt}
		Initial & Final & \multicolumn{2}{c}{$\Omega_{ccb}$} & \multicolumn{2}{c}{$\Omega_{bbc}$}
		\\
		\noalign{\vskip 2pt}
		\cline{3-4}
		\cline{5-6}
		\noalign{\vskip 2pt}
		State & State & $E_{\gamma}$(MeV) & $\Gamma$(KeV) & $E_{\gamma}$(MeV) & $\Gamma$(KeV) 
		\\
		\hline
		\noalign{\vskip 2pt}
		$2S1p\left(1/2^{-}\right)$ & $1S1s\left(1/2^{+}\right)$ & 689.77 & 1.228 & 713.34 & 0.078 
		\\
		 & $1S2s\left(1/2^{+}\right)$ & 133.48 & 0.022 & 155.76 & 0.002 
		\\
		 & $2S1s\left(1/2^{+}\right)$ & 406.74 & 0.293 & 419.49 & 0.023 
		\\
		 & $1S1p'\left(1/2^{-}\right)$ & 268.89 & 0.017 & 270.90 & 6.661 
		\\
		 & $1S1p'\left(3/2^{-}\right)$ & 246.09 & 0.019 & 260.24 & 15.195 
		\\
		 & $1P1s\left(1/2^{-}\right)$ & 461.23 & 0.817 & 490.82 & 188.222 
		\\
		 & $2P1s\left(1/2^{-}\right)$ & 265.16 & 0.163 & 269.51 & 39.397 
		\\
		 & $1D1s\left(1/2^{+}\right)$ & 319.97 & $\sim$0 & 328.49 & 7.372 
		\\
		\hline
		\noalign{\vskip 2pt}
		$2S1p'\left(1/2^{-}\right)$ & $1S1p\left(1/2^{-}\right)$ & 324.83 & 0.028 & 338.46 & 12.751 
		\\
		 & $1S1p'\left(3/2^{-}\right)$ & 284.43 & 0.029 & 301.76 & 13.547 
		\\
		 & $2S1p\left(1/2^{-}\right)$ & 28.47 & $\sim$0 & 34.48 & 0.017 
		\\
		 & $1P1s\left(3/2^{-}\right)$ & 469.34 & 1.284 & 494.85 & 16.531 
		\\
		 & $2P1s\left(3/2^{-}\right)$ & 273.72 & 0.268 & 273.70 & 3.545 
		\\
		\hline
		\noalign{\vskip 2pt}
		$2S1p\left(3/2^{-}\right)$ & $1S1s\left(1/2^{+}\right)$ & 726.93 & 1.031 & 747.95 & 0.135 
		\\
		 & $1S2s\left(1/2^{+}\right)$ & 173.28 & 0.035 & 192.11 & 0.006 
		\\
		 & $2S1s\left(1/2^{+}\right)$ & 445.24 & 0.280 & 455.01 & 0.045 
		\\
		 & $1S1p'\left(1/2^{-}\right)$ & 308.04 & 0.018 & 306.88 & 6.443 
		\\
		 & $1S1p'\left(3/2^{-}\right)$ & 285.36 & 0.009 & 296.26 & 5.953 
		\\
		 & $1S1p\left(5/2^{-}\right)$ & 269.68 & $\sim$0 & 280.36 & 12.241 
		\\
		 & $1P1s\left(1/2^{-}\right)$ & 499.47 & 0.752 & 526.12 & 152.526 
		\\
		 & $2P1s\left(1/2^{-}\right)$ & 304.33 & 0.181 & 305.49 & 38.471 
		\\
		 & $1P1p'\left(1/2^{+}\right)$ & 77.27 & 0.026 & 81.91 & 0.004 
		\\
		 & $1P1p'\left(3/2^{+}\right)$ & 70.68 & 0.012 & 74.43 & 0.002 
		\\
		 & $1D1s\left(1/2^{+}\right)$ & 358.88 & $\sim$0 & 364.29 & 6.557 
		\\
		\hline
		\noalign{\vskip 2pt}
		$2S1p'\left(3/2^{-}\right)$ & $1S1p\left(1/2^{-}\right)$ & 346.16 & 0.024 & 348.75 & 18.534 
		\\
		 & $1S1p\left(3/2^{-}\right)$ & 305.87 & 0.011 & 312.08 & 7.958 
		\\
		 & $2S1p\left(1/2^{-}\right)$ & 50.57 & $\sim$0 & 45.04 & 0.052 
		\\
		 & $1P1s\left(3/2^{-}\right)$ & 490.31 & 0.430 & 505.00 & 8.393 
		\\
		 & $2P1s\left(3/2^{-}\right)$ & 295.18 & 0.099 & 284.05 & 1.903 
		\\
		\hline
		\noalign{\vskip 2pt}
		$2S1p\left(5/2^{-}\right)$ & $1S1s\left(3/2^{+}\right)$ & 695.29 & $\sim$0 & 719.28 & 93.975 
		\\
		 & $1S2s\left(3/2^{+}\right)$ & 171.24 & $\sim$0 & 184.69 & 4.376 
		\\
		 & $2S1s\left(3/2^{+}\right)$ & 415.11 & $\sim$0 & 426.79 & 28.582 
		\\
		 & $1S1p\left(3/2^{-}\right)$ & 321.27 & $\sim$0 & 327.63 & 15.702 
		\\
		 & $1P1s\left(3/2^{-}\right)$ & 505.38 & $\sim$0 & 520.29 & 45.517 
		\\
		 & $2P1s\left(3/2^{-}\right)$ & 310.62 & $\sim$0 & 299.64 & 11.164 
		\\
		 & $1D1s\left(3/2^{+}\right)$ & 377.98 & $\sim$0 & 397.03 & 0.230 
		\\
		 & $1D1s'\left(3/2^{+}\right)$ & 372.31 & $\sim$0 & 381.01 & 0.651 
		\\
		\hline
	\end{tabular*}
\end{table}

\begin{table}[!htbp]
	\caption{\label{tab:11} Radiative decays of $1S1d$ state.}
	\centering
	\begin{tabular*}{\textwidth}{@{\extracolsep{\fill}}cccccc}
		\hline
		\noalign{\vskip 2pt}
		Initial & Final & \multicolumn{2}{c}{$\Omega_{ccb}$} & \multicolumn{2}{c}{$\Omega_{bbc}$}
		\\
		\noalign{\vskip 2pt}
		\cline{3-4}
		\cline{5-6}
		\noalign{\vskip 2pt}
		State & State & $E_{\gamma}$(MeV) & $\Gamma$(KeV) & $E_{\gamma}$(MeV) & $\Gamma$(KeV) 
		\\
		\hline
		\noalign{\vskip 2pt}
		$1S1d\left(1/2^{+}\right)$ & $1P1p\left(1/2^{+}\right)$ & 50.77 & 0.179 & 80.75 & 1.006 
		\\
		 & $1P1p\left(3/2^{+}\right)$ & 42.10 & 0.154 & 51.81 & 0.397 
		\\
		 & $1D1s\left(3/2^{+}\right)$ & 314.46 & 29.986 & 350.34 & 23.058 
		\\
		 & $1D1s'\left(3/2^{+}\right)$ & 308.74 & 6.103 & 334.25 & 57.360 
		\\
		\hline
		\noalign{\vskip 2pt}
		$1S1d\left(3/2^{+}\right)$ & $1S1p\left(1/2^{-}\right)$ & 303.61 & 0.065 & 307.93 & 26.865 
		\\
		 & $1S1p\left(3/2^{-}\right)$ & 263.11 & 0.026 & 271.12 & 10.262 
		\\
		 & $1P1s\left(3/2^{-}\right)$ & 448.50 & 0.436 & 464.73 & 1.627 
		\\
		 & $2P1s\left(3/2^{-}\right)$ & 252.37 & 0.083 & 242.99 & 0.298 
		\\
		 & $1P1p\left(1/2^{+}\right)$ & 56.71 & 0.296 & 71.06 & 0.162 
		\\
		 & $1P1p\left(3/2^{+}\right)$ & 48.05 & 0.108 & 42.10 & 0.020 
		\\
		 & $1D1s\left(3/2^{+}\right)$ & 320.21 & 7.925 & 340.86 & 1.538 
		\\
		 & $1D1s'\left(3/2^{+}\right)$ & 314.49 & 2.559 & 324.77 & 2.792 
		\\
		\hline
		\noalign{\vskip 2pt}
		$1S1d'\left(3/2^{+}\right)$ & $1S1p'\left(1/2^{-}\right)$ & 281.71 & 0.199 & 290.48 & 4.653
		\\
		 & $1S1p'\left(3/2^{-}\right)$ & 258.96 & 0.095 & 279.85 & 5.195 
		\\
		 & $1P1s\left(1/2^{-}\right)$ & 473.75 & 1.132 & 510.00 & 0.321 
		\\
		 & $2P1s\left(1/2^{-}\right)$ & 277.99 & 0.247 & 289.09 & 0.077 
		\\
		 & $1P1p\left(5/2^{+}\right)$ & 47.16 & 0.017 & 72.76 & 0.219 
		\\
		 & $1D1s\left(1/2^{+}\right)$ & 332.72 & 3.377 & 347.98 & 33.755 
		\\
		 & $1D1s\left(5/2^{+}\right)$ & 317.30 & 2.899 & 352.44 & 9.484 
		\\
		 & $1D1s'\left(5/2^{+}\right)$ & 316.42 & 0.383 & 329.72 & 10.445 
		\\
		\hline
		\noalign{\vskip 2pt}
		$1S1d\left(5/2^{+}\right)$ & $1S1p\left(3/2^{-}\right)$ & 272.69 & 0.023 & 276.32 & 19.889 
		\\
		 & $1P1s\left(3/2^{-}\right)$ & 457.87 & 0.366 & 469.83 & 0.688 
		\\
		& $2P1s\left(3/2^{-}\right)$ & 261.96 & 0.074 & 248.20 & 0.130 
		\\
		 & $1P1p\left(3/2^{+}\right)$ & 57.88 & 0.138 & 47.39 & 0.060 
		\\
		 & $1D1s\left(3/2^{+}\right)$ & 329.73 & 6.260 & 346.03 & 3.402 
		\\
		 & $1D1s'\left(3/2^{+}\right)$ & 324.02 & 2.027 & 329.93 & 5.922 
		\\
		 & $1D1s\left(7/2^{+}\right)$ & 314.49 & 2.065 & 333.91 & 3.163 
		\\
		\hline
		\noalign{\vskip 2pt}
		$1S1d\left(5/2'^{+}\right)$ & $1S1p'\left(3/2^{-}\right)$ & 262.57 & 0.072 & 291.44 & 2.079 
		\\
		 & $1P1p\left(5/2^{+}\right)$ & 50.86 & 0.011 & 84.56 & 0.077 
		\\
		 & $1D1s\left(5/2^{+}\right)$ & 320.88 & 1.510 & 363.96 & 2.478 
		\\
		 & $1D1s'\left(5/2^{+}\right)$ & 320.00 & 0.199 & 341.27 & 2.703 
		\\
		\hline
		\noalign{\vskip 2pt}
		$1S1d\left(7/2^{+}\right)$ & $1S1p\left(5/2^{-}\right)$ & 251.74 & $\sim$0 & 259.90 & 12.317 
		\\
		 & $1P1p\left(5/2^{+}\right)$ & 55.87 & 0.035 & 68.67 & 0.101 
		\\
		 & $1D1s\left(5/2^{+}\right)$ & 325.73 & 1.861 & 348.45 & 1.712 
		\\
		 & $1D1s'\left(5/2^{+}\right)$ & 324.85 & 0.359 & 325.72 & 2.610 
		\\
		\hline
	\end{tabular*}
\end{table}

\begin{table}
	\caption{\label{tab:12} Radiative decays of $1P1s, 1P2s, 2P1s$ and $1P1p$ states.}
	\centering
	\begin{tabular*}{\textwidth}{@{\extracolsep{\fill}}cccccc}
		\hline
		\noalign{\vskip 2pt}
		Initial & Final & \multicolumn{2}{c}{$\Omega_{ccb}$} & \multicolumn{2}{c}{$\Omega_{bbc}$}
		\\
		\noalign{\vskip 2pt}
		\cline{3-4}
		\cline{5-6}
		\noalign{\vskip 2pt}
		State & State & $E_{\gamma}$(MeV) & $\Gamma$(KeV) & $E_{\gamma}$(MeV) & $\Gamma$(KeV) 
		\\
		\hline
		\noalign{\vskip 2pt}
		$1P1s\left(3/2^{-}\right)$  & $1S1s\left(1/2^{+}\right)$ & 261.38 & 1.054 & 262.08 & 0.792 
		\\
		 & $1P1s\left(1/2^{-}\right)$ & 19.94 & 0.002 & 30.32 & 0.010 
		\\
		\hline
		\noalign{\vskip 2pt}
		$1P2s\left(3/2^{-}\right)$ & $1S1s\left(1/2^{+}\right)$ & 787.71 & 1.914 & 801.99 & 0.886 
		\\
		 & $1S2s\left(1/2^{+}\right)$ & 238.34 & 3.449 & 248.84 & 2.637 
		\\
		 & $2S1s\left(1/2^{+}\right)$ & 508.19 & 0.866 & 510.47 & 0.739 
		\\
		 & $1S1p'\left(1/2^{-}\right)$ & 372.06 & 0.146 & 363.06 & 0.539 
		\\
		 & $1S1p'\left(3/2^{-}\right)$ & 349.55 & 0.077 & 352.49 & 0.412 
		\\
		 & $1P1s\left(1/2^{-}\right)$ & 562.00 & 0.208 & 581.24 & 0.812 
		\\
		 & $2P1s\left(1/2^{-}\right)$ & 368.38 & 0.023 & 361.68 & 0.040 
		\\
		 & $1P1p'\left(1/2^{+}\right)$ & 143.08 & 2.082 & 139.18 & 1.343 
		\\
		 & $1P1p'\left(3/2^{+}\right)$ & 136.53 & 1.089 & 131.74 & 0.676 
		\\
		\hline
		\noalign{\vskip 2pt}
		$2P1s\left(1/2^{-}\right)$ & $1S1p\left(1/2^{-}\right)$ & 33.24 & $\sim$0 & 36.19 & 0.012 
		\\
		 & $1P1s\left(3/2^{-}\right)$ & 182.99 & 2.335 & 196.74 & 5.148 
		\\
		\hline
		\noalign{\vskip 2pt}
		$2P1s\left(3/2^{-}\right)$ & $1S1s\left(1/2^{+}\right)$ & 457.12 & 4.839 & 483.34 & 4.382 
		\\
		 & $2S1s\left(1/2^{+}\right)$ & 165.53 & 0.239 & 183.33 & 0.283 
		\\
		 & $1P1s\left(1/2^{-}\right)$ & 221.66 & 2.048 & 256.16 & 5.539 
		\\
		 & $2P1s\left(1/2^{-}\right)$ & 19.67 & 0.001 & 30.20 & 0.010 
		\\
		\hline
		\noalign{\vskip 2pt}
		$1P1p\left(1/2^{+}\right)$  & $1S1p'\left(1/2^{-}\right)$ & 219.93 & 0.656 & 204.14 & 0.322 
		\\
		 & $1S1p'\left(3/2^{-}\right)$ & 197.00 & 0.717 & 193.42 & 0.377 
		\\
		 & $1P1s\left(1/2^{-}\right)$ & 413.42 & 114.924 & 425.34 & 90.501 
		\\
		 & $2P1s\left(1/2^{-}\right)$ & 216.18 & 17.063 & 202.74 & 11.604 
		\\
		 & $1D1s\left(1/2^{+}\right)$ & 271.31 & 1.099 & 262.07 & 1.629 
		\\
		\hline
		\noalign{\vskip 2pt}
		$1P1p\left(3/2^{+}\right)$ & $1S1p'\left(1/2^{-}\right)$ & 228.41 & 0.548 & 232.69 & 0.344 
		\\
		 & $1S1p'\left(3/2^{-}\right)$ & 205.51 & 0.243 & 222.01 & 0.165 
		\\
		 & $1P1s\left(1/2^{-}\right)$ & 421.70 & 90.903 & 453.35 & 78.644 
		\\
		 & $2P1s\left(1/2^{-}\right)$ & 224.67 & 14.309 & 231.30 & 12.523 
		\\
		 & $1D1s\left(1/2^{+}\right)$ & 279.74 & 0.898 & 290.48 & 1.602 
		\\
		 & $1D1s\left(5/2^{+}\right)$ & 264.23 & 1.736 & 294.97 & 1.589 
		\\
		 & $1D1s'\left(5/2^{+}\right)$ & 263.34 & 0.334 & 272.13 & 2.301 
		\\
		\hline
		\noalign{\vskip 2pt}
		$1P1p\left(5/2^{+}\right)$  & $1S1p\left(3/2^{-}\right)$ & 223.74 & 1.019 & 216.22 & 0.113 
		\\
		 & $1P1s\left(3/2^{-}\right)$ & 410.02 & 38.189 & 410.75 & 30.793 
		\\
		 & $2P1s\left(3/2^{-}\right)$ & 212.95 & 5.548 & 187.96 & 3.469 
		\\
		 & $1D1s\left(3/2^{+}\right)$ & 281.11 & 0.925 & 286.30 & 1.253 
		\\
		 & $1D1s'\left(3/2^{+}\right)$ & 275.38 & 0.496 & 270.12 & 1.223 
		\\
		\hline
	\end{tabular*}
\end{table}

\begin{table}
	\caption{\label{tab:13} Radiative decays of $1D1s, 1D1p$ and $1F1s$ states.}
	\centering
	\begin{tabular*}{\textwidth}{@{\extracolsep{\fill}}cccccc}
		\hline
		\noalign{\vskip 2pt}
		Initial & Final & \multicolumn{2}{c}{$\Omega_{ccb}$} & \multicolumn{2}{c}{$\Omega_{bbc}$}
		\\
		\noalign{\vskip 2pt}
		\cline{3-4}
		\cline{5-6}
		\noalign{\vskip 2pt}
		State & State & $E_{\gamma}$(MeV) & $\Gamma$(KeV) & $E_{\gamma}$(MeV) & $\Gamma$(KeV) 
		\\
		\hline
		\noalign{\vskip 2pt}
		$1D1s\left(3/2^{+}\right)$ & $1P1s\left(1/2^{-}\right)$ & 153.02 & 0.011 & 157.58 & 0.049 
		\\
		\hline
		\noalign{\vskip 2pt}
		$1D1s'\left(3/2^{+}\right)$ & $1P1s\left(1/2^{-}\right)$ & 158.85 & 0.066 & 173.91 & 0.021 
		\\
		\hline
		\noalign{\vskip 2pt}
		$1D1s\left(5/2^{+}\right)$ & $1P1s\left(3/2^{-}\right)$ & 143.04 & 0.003 & 132.54 & 0.024 
		\\
		\hline
		\noalign{\vskip 2pt}
		$1D1s'\left(5/2^{+}\right)$ & $1P1s\left(3/2^{-}\right)$ & 143.94 & 0.084 & 155.65 & 0.054 
		\\
		\hline
		\noalign{\vskip 2pt}
		$1D1p\left(1/2^{-}\right)$ & $1S1p\left(1/2^{-}\right)$ & 326.07 & 0.010 & 360.72 & 22.911 
		\\
		 & $2S1p\left(1/2^{-}\right)$ & 67.06 & $\sim$0 & 57.32 & 0.122 
		\\
		 & $1S1p\left(3/2^{-}\right)$ & 321.86 & 0.012 & 324.08 & 24.844 
		\\
		 & $1P1s\left(3/2^{-}\right)$ & 505.94 & 0.451 & 516.80 & 20.957 
		\\
		 & $2P1s\left(3/2^{-}\right)$ & 311.19 & 0.112 & 296.08 & 5.054 
		\\
		\hline
		\noalign{\vskip 2pt}
		$1D1p\left(3/2^{-}\right)$  & $1S1s\left(1/2^{+}\right)$ & 784.77 & 0.352 & 798.66 & $\sim$0 
		\\
		 & $1S2s\left(1/2^{+}\right)$ & 235.19 & 0.024 & 245.34 & $\sim$0 
		\\
		 & $2S1s\left(1/2^{+}\right)$ & 505.15 & 0.114 & 507.05 & $\sim$0 
		\\
		 & $1S1p'\left(1/2^{-}\right)$ & 368.96 & 0.008 & 359.59 & 0.248 
		\\
		 & $1S1p'\left(3/2^{-}\right)$ & 346.45 & 0.004 & 349.02 & 0.231 
		\\
		 & $1S1d\left(1/2^{+}\right)$ & 102.81 & $\sim$0 & 77.97 & 0.015 
		\\
		 & $1S1d\left(5/2^{+}\right)$ & 87.12 & $\sim$0 & 82.38 & 0.011 
		\\
		 & $1P1s\left(1/2^{-}\right)$ & 558.98 & 0.293 & 577.83 & 0.108 
		\\
		 & $2P1s\left(1/2^{-}\right)$ & 365.28 & 0.087 & 358.21 & 0.033 
		\\
		 & $1P1p'\left(1/2^{+}\right)$ & 139.89 & 0.044 & 135.65 & $\sim$0 
		\\
		 & $1P1p'\left(3/2^{+}\right)$ & 133.35 & 0.023 & 128.19 & $\sim$0 
		\\
		 & $1P1p\left(5/2^{+}\right)$ & 136.88 & $\sim$0 & 143.17 & 0.061 
		\\
		 & $1D1s\left(1/2^{+}\right)$ & 419.43 & $\sim$0 & 416.75 & 1.545 
		\\
		 & $1D1s\left(5/2^{+}\right)$ & 404.18 & $\sim$0 & 421.18 & 0.971 
		\\
		 & $1D1s'\left(5/2^{+}\right)$ & 403.30 & $\sim$0 & 398.60 & 1.126 
		\\
		\hline
		\noalign{\vskip 2pt}
		$1D1p'\left(3/2^{-}\right)$ & $1S1p'\left(1/2^{-}\right)$ & 372.42 & $\sim$0 & 362.84 & 0.891 
		\\
		 & $1S1p'\left(3/2^{-}\right)$ & 349.91 & $\sim$0 & 352.27 & 0.821 
		\\
		 & $1S1p\left(5/2^{-}\right)$ & 334.36 & $\sim$0 & 336.44 & 2.052 
		\\
		 & $1S1d\left(1/2^{+}\right)$ & 106.38 & $\sim$0 & 81.29 & 0.018 
		\\
		 & $1P1s\left(1/2^{-}\right)$ & 562.36 & $\sim$0 & 581.02 & 16.503 
		\\
		 & $2P1s\left(1/2^{-}\right)$ & 368.74 & $\sim$0 & 361.46 & 5.267 
		\\
		 & $1D1s\left(1/2^{+}\right)$ & 422.86 & $\sim$0 & 419.97 & 0.769 
		\\
		 & $1D1s\left(5/2^{+}\right)$ & 407.62 & $\sim$0 & 424.41 & 0.012 
		\\
		 & $1D1s'\left(5/2^{+}\right)$ & 406.74 & $\sim$0 & 401.83 & 0.019 
		\\
		\hline
		\noalign{\vskip 2pt}
		$1D1p\left(5/2^{-}\right)$  & $1S1d'\left(3/2^{+}\right)$ & 96.54 & 0.005 & 66.91 & 0.292 
		\\
		 & $1S1d'\left(5/2^{+}\right)$ & 92.85 & 0.003 & 55.09 & 0.061 
		\\
		 & $1P1p\left(3/2^{+}\right)$ & 150.69 & 0.095 & 125.52 & 0.032 
		\\
		 & $1D1s\left(3/2^{+}\right)$ & 419.58 & 0.021 & 422.18 & 9.146 
		\\
		 & $1D1s'\left(3/2^{+}\right)$ & 413.93 & 0.111 & 406.19 & 20.404 
		\\
		 & $1F1s\left(7/2^{-}\right)$ & 303.75 & 0.074 & 302.12 & 4.179 
		\\
		\hline
		\noalign{\vskip 2pt}
		$1D1p'\left(5/2^{-}\right)$  & $1S1p\left(3/2^{-}\right)$ & 366.43 & $\sim$0 & 361.79 & 2.330 
		\\
		 & $2S1p\left(3/2^{-}\right)$ & 72.91 & $\sim$0 & 59.26 & 0.012 
		\\
		 & $1S1d'\left(3/2^{+}\right)$ & 99.90 & 0.005 & 75.99 & 0.015 
		\\
		 & $1P1s\left(3/2^{-}\right)$ & 549.51 & $\sim$0 & 553.89 & 0.089 
		\\
		 & $2P1s\left(3/2^{-}\right)$ & 355.82 & $\sim$0 & 333.89 & 0.025 
		\\
		 & $1P1p\left(3/2^{+}\right)$ & 154.04 & 0.102 & 134.56 & 0.103 
		\\
		 & $1D1s\left(3/2^{+}\right)$ & 422.82 & 0.022 & 430.99 & 0.848 
		\\
		 & $1D1s'\left(3/2^{+}\right)$ & 417.17 & 0.113 & 415.02 & 1.603 
		\\
		 & $1D1s\left(7/2^{+}\right)$ & 407.76 & $\sim$0 & 418.97 & 0.746 
		\\
		 & $1F1s\left(7/2^{-}\right)$ & 307.04 & 0.076 & 311.02 & $\sim$0 
		\\
		\hline
		\noalign{\vskip 2pt}
		$1D1p\left(7/2^{-}\right)$  & $1P1p\left(5/2^{+}\right)$ & 144.82 & 0.157 & 142.61 & 0.051 
		\\
		 & $1D1s\left(5/2^{+}\right)$ & 411.86 & 0.002 & 420.64 & 0.011 
		\\
		 & $1D1s'\left(5/2^{+}\right)$ & 410.99 & 0.067 & 398.05 & 0.014 
		\\
		 & $1F1s\left(5/2^{-}\right)$ & 308.27 & 0.058 & 277.02 & 0.034 
		\\
		\hline
		\noalign{\vskip 2pt}
		$1F1s\left(5/2^{-}\right)$  & $1D1s\left(3/2^{+}\right)$ & 117.17 & 0.004 & 152.01 & 0.034 
		\\
		\hline
		\noalign{\vskip 2pt}
		$1F1s\left(7/2^{-}\right)$  & $1D1s\left(5/2^{+}\right)$ & 110.48 & 0.001 & 118.29 & 0.010 
		\\
		\hline
		\end{tabular*}
\end{table}

We have evaluated the radiative decay widths of the $\Omega_{ccb}$ and $\Omega_{bbc}$ baryons using the wave functions obtained within our model, limiting our analysis to the electric dipole (E1) and magnetic dipole (M1) contributions. The calculated decay widths and corresponding photon energies for baryon states containing  $S-$ wave diquarks presented in Tables \ref{tab:9}, \ref{tab:10} and \ref{tab:11}, while the results for states with P-wave diquarks are listed in Table \ref{tab:12} The radiative transitions involving baryon states with $D$ and $F-$ wave diquarks are summarized in Table \ref{tab:13}. In all cases, the decay widths $\Gamma$ are expressed in KeV, whereas the photon energies $E_{\gamma}$ are given in MeV. For clarity, only transitions with $\Gamma \geq 0.01$ KeV are included. The ground-state transition $1S1s(3/2^{+}) \rightarrow 1S1s(1/2^{+})$ has a radiative decay width of $0.015$ KeV for both the $\Omega_{ccb}$ and $\Omega_{bbc}$ baryons. More generally, spin-flip transitions $(N_{d}L_{d}n_{q}l_{q}(S) \rightarrow \gamma N_{d}L_{d}n_{q}l_{q}(S'))$, which occur between states with the same radial and orbital quantum numbers, are consistently suppressed due to the heavy-quark spin symmetry and the relatively small hyper-fine mass splittings between the corresponding states. This suppression persists throughout the excited-state spectrum, including the transitions $2S1s(3/2^{+}) \rightarrow 2S1s(1/2^{+})$, $1S1p'(3/2^{-}) \rightarrow 1S1p(1/2^{-})$, $2S1p'(3/2^{-}) \rightarrow 2S1p(1/2^{-})$, $1P1s(3/2^{-}) \rightarrow 1P1s(1/2^{-})$, and $2P1s(3/2^{-}) \rightarrow 2P1s(1/2^{-})$, all of which exhibit very small radiative decay widths. A similar suppression is observed for transitions occurring within the same fine-structure multiplet, such as $1S1p'(1/2^{-}) \rightarrow 1S1p(1/2^{-})$ and  $2S1p'(1/2^{-}) \rightarrow 2S1p(1/2^{-})$. These transitions are also governed by heavy-quark spin symmetry and involve only spin-changing electromagnetic operators while preserving the radial and orbital quantum numbers. Consequently, the corresponding transition matrix elements are strongly suppressed, resulting in very small radiative decay widths. The most favored radiative transitions are those involving changes in the radial quantum number, orbital quantum number, or both, particularly when accompanied by sufficiently large photon energies. Typical examples include $2S1s \rightarrow 1S1s$, $1S1p, 1S1p' \rightarrow 1P1s$, $2S1p, 2S1p' \rightarrow 1P1s, 2P1s$ and  $1S1d, 1S1d' \rightarrow 1S1p, 1S1p', 1D1s, 1D1s'$ for $S-$ diquark baryon states. For baryons containing $P-$ wave diquarks, the dominant decay modes include $1P1s \rightarrow 1S1s$, $2P1s \rightarrow 1P1s$ and $1P1p \rightarrow 1P1s, 2P1s$. Similarly, for baryons with $D-$ wave diquarks, transitions such as $1D1s \rightarrow 1P1s$ and $1D1p \rightarrow 1S1p, 1P1s, 1D1s, 1D1s'$ constitute the primary decay channels. The comparatively large decay widths of these transitions indicate that changes in the radial and/or orbital structure of the wave functions lead to significantly larger overlap integrals, thereby enhancing the corresponding transition amplitudes. The calculated radiative transitions also reveal a clear distinction between excitations associated with relative quark-diquark motion and those involving internal diquark excitations. States corresponding to excitations of the quark-diquark orbital quantum number $N_{l}$, such as $1S1p, 2S1p, 1P1p$ and $1D1p$ exhibit substantially richer radiative decay spectra. In contrast, states associated with excitations of the diquark orbital quantum number $N_{d}$, such as $1P1s, 2P1s$ and $1D1s$, possess comparatively fewer radiative decay channels. Rather than serving as unstable excited states, these diquark-excited configurations primarily act as intermediate states in the radiative cascade, through which higher excited baryons de-excite. Consequently, most excited configurations first populate the $1P1s, 2P1s$ and $1D1s$ multiplets before undergoing subsequent radiative transitions to the ground state. For instance, the the $1S1p$ and $2S1p$ multiplet predominantly decay into $1P1s$ states, the $1P1p$ multiplet also favors transitions to the $1P1s$ states, while the $1S1d$ and $1D1p$ decay preferentially to the $1D1s$ and $1P1s$ multiplets. The radiative decay widths do not exhibit a simple monotonic dependence on the heavy-quark content. Although the decay channels and corresponding photon energies are broadly similar for both baryons, several transitions involving orbital excitations are predicted to be considerably stronger in the $\Omega_{bbc}$ system. This behavior may be attributed to the more compact spatial distribution of the $bb$ diquark, which results in a stronger localization of the baryon wave functions and consequently enhances the corresponding radial overlap integrals in the transition amplitudes. As a result, many transitions that are strongly suppressed in the $\Omega_{ccb}$ baryon, where the corresponding widths are either numerically negligible or effectively forbidden due to cancellations arising from the orthogonality of the radial wave functions, become experimentally significant in the case of $\Omega_{bbc}$ baryon. These differences between the two baryon may provide valuable insight into the internal dynamics of the heavy diquark and may serve as sensitive probes of the underlying quark-diquark structure. Precise measurements of radiative branching ratios could provide important constraints on assignment of excited states, reveal the ordering of excited levels, and distinguish between different excitation modes. The present results suggest that radiative decay widths are a powerful tool to investigating the internal dynamics of triply heavy baryons and can provide clear experimental signatures for identifying and distinguishing different quark-diquark and diquark excitation modes.

\section{\label{sec:Conclusion} Conclusion}

In this study, we have applied a relativistic screened potential model within the quark–diquark framework to systematically investigate the mass spectra of the $\Omega_{ccb}$ and $\Omega_{bbc}$ baryons. We have analyzed the ground ($S$-wave) as well as excited ($P$- and $D$-wave) states across various quark–diquark configurations. The calculated mass spectra for $S$-, $P$-, and $D$-wave states provide consistent and physically motivated insights into the excitation dynamics of triply heavy baryons. Our findings indicate that diquark excitations, contrary to conventional expectations, often lie lower in energy than those involving the quark–diquark relative motion, suggesting a more accessible excitation spectrum for the heavy diquark subsystem. Moreover, the quark–diquark approximation inherently reduces the number of allowed excited states due to symmetry constraints imposed by the Pauli exclusion principle. This feature offers a natural explanation for the so-called missing resonance problem, supporting the idea that many expected states may be suppressed or absent within this framework. Alongside the spectroscopy, we have also carried out a detailed analysis of the radiative decays of these states. The results suggest that spin–flip transitions within the same orbital multiplet are strongly suppressed, whereas transitions involving changes in orbital or radial quantum numbers dominate and generally exhibit much larger widths. The comparison between $\Omega_{ccb}$ and $\Omega_{bbc}$ further shows how differences in their internal dynamics significantly influence the decay widths. Overall, our findings reinforce the validity of the quark–diquark picture for modeling triply heavy baryons and highlight the importance of internal diquark excitations in understanding their spectra and decays.

\section*{Acknowledgements}
One of the authors (CAB) is grateful to Manipal Academy of Higher Education (MAHE) for the financial support under the Dr. T.M.A. Pai Scholarship Program.

\section*{Funding}
This research did not receive funding from any funding
agency.

\section*{Data Availability}
Data sharing not applicable to this article as no
datasets were generated or analysed during the current study.

\section*{Code Availability}
Code/Software sharing not applicable
to this article as no code/software was generated or analysed during
the current study.

\nocite{*}

\bibliography{bbcandccbBaryonsArxiv}

\end{document}